\documentclass[fleqn,usenatbib]{mnras}
\usepackage{fix-cm}
\usepackage{newtxtext,newtxmath}

\usepackage[T1]{fontenc}

\DeclareRobustCommand{\VAN}[3]{#2}
\let\VANthebibliography\thebibliography
\def\thebibliography{\DeclareRobustCommand{\VAN}[3]{##3}\VANthebibliography}

\usepackage{graphicx}	
\usepackage{amsmath}	
\usepackage{subcaption}
\usepackage{multirow}
\usepackage{float}
\usepackage{gensymb}
\usepackage[pagewise]{lineno}

\title[Compton-thick AGN: A Multi-band Perspective]{Compton-thick AGN Characterisation in a Multi-wavelength Context: Insights from the 70-Month \textit{SWIFT}/BAT Catalogue}

\author[M. L. H. Musa et al. 2026]{
M. L. H. Musa,$^{1,2}$
Z. Z. Abidin,$^{1,3}$\thanks{E-mail: zzaa@um.edu.my}
M. Imanishi$^{4,5}$
Y. Hagiwara,$^{6}$
A. Annuar,$^{7}$
\\
$^{1}$Radio Cosmology Research Lab, Centre for Astronomy and Astrophysics Research, Faculty of Science, Universiti Malaya, 50603, Kuala Lumpur, Malaysia\\
$^{2}$Centre of Foundation, Language and Malaysian Studies, Universiti Malaya-Wales, 50480, Kuala Lumpur, Malaysia\\
$^{3}$National Centre for Particle Physics (NCPP), Universiti Malaya, 50603, Kuala Lumpur, Malaysia\\
$^{4}$National Astronomical Observatory of Japan, National Institutes of Natural Sciences (NINS), 2-21-1 Osawa, Mitaka, Tokyo 181-8588, Japan\\
$^{5}$Department of Astronomical Science, SOKENDAI (The Graduate University of Advanced Studies), 2-21-1 Osawa, Mitaka, Tokyo 181-8588, Japan\\
$^{6}$Natural Science Laboratory, Toyo University, 5-28-20, Hakusan, Bunkyo-ku, Tokyo 112-8606, Japan\\
$^{7}$Department of Applied Physics, Faculty of Science and Technology, Universiti Kebangsaan Malaysia, 43600 UKM Bangi, Selangor, Malaysia\\
}

\date{Accepted XXX. Received YYY; in original form ZZZ}

\pubyear{\the\year{}}

\begin{document}
\label{firstpage}
\pagerange{\pageref{firstpage}--\pageref{lastpage}}
\maketitle

\begin{abstract}
{
We analyse Compton-thick active galactic nuclei (CT AGNs), a heavily obscured subclass that challenges traditional X-ray diagnostics. Using 243 sources from the 70-Month \textit{SWIFT}/BAT catalogue (26 CT, 217 non-CT), we investigate their properties across radio, infrared (IR), optical, and X-ray bands. VLASS data reveals slightly higher 2--3~GHz mean luminosities in CT AGNs, suggesting active cores attenuated by circumnuclear absorption. Mid-IR diagnostics show redder $W3-W4$ colours in CT AGNs, tracing cooler dust, with significant scatter likely driven by host-galaxy dilution. Most CT AGNs fall outside standard WISE selection wedges, highlighting mid-IR selection limitations. BPT diagnostics show that CT AGNs primarily occupy Seyfert regions, indicating isotropic narrow-line properties. CT AGNs favour significantly higher Eddington ratios ($\lambda_{\text{Edd}}$), supporting radiation-driven unification where intense accretion maintains high-column density. We also observe a moderate anti-correlation between [NII]/H$\alpha$ and $\lambda_{\text{Edd}}$. Principal component analysis identifies ionizing power and the accretion-obscuration link as primary variance drivers, though both populations overlap significantly in the PC1--PC2 plane. Machine learning achieved high recall (0.80) using intrinsic X-ray luminosity, [OI]$\lambda$6300 and H$\alpha$ luminosities, and $W2-W3$ colour. This demonstrates the potential for multi-wavelength signatures to verify CT candidates in future deep surveys where X-ray data is limited. Overall, our findings suggest CT AGNs are driven by high obscuration and accretion rates rather than a simple orientation effect.}

\end{abstract}



\section{Introduction}

Active galactic nuclei (AGNs) are among the most energetic processes in the universe, emitting intense radiation observable across the electromagnetic spectrum. These emissions originate from various processes occurring in the central engine, which are believed to be powered by a supermassive black hole (SMBH). The SMBH actively accretes material at the centre, potentially influencing the evolution of the host galaxy through processes such as AGN feedback, making it a crucial aspect of galaxy evolution studies \cite[e.g.][]{heckman2014coevolution,gatti2015physical, hlavacek2024agn, arjona2024role}. The diverse processes within the central region of AGN take place in distinct components, described by the AGN unification model \citep {antonucci1993unified}. Over the past few decades, various observational techniques, spanning from radio to gamma rays, have been employed to identify these powerful sources \cite[e.g.][]{haas2004mid,padovani2015radio,tadhunter2016radio,zaw2019uniformly,radcliffe2021nowhere}. This broad wavelength coverage allows astronomers to obtain a comprehensive understanding of AGN properties, as different wavelengths provide unique insights into their physical characteristics. For instance, radio emissions are often associated with radio jets and lobes, typically observed in quasars. Infrared (IR) observations trace the obscuring material, composed of dust and gas, while optical and ultraviolet (UV) emissions are linked to the accretion disc. High-energy emissions, such as X-rays, result from the inverse-Compton scattering process occurring in the putative corona, while gamma-ray emissions are typically associated with strong emitters like non-thermal AGN jets or lobes.

AGN can be categorised into different types, based on the physical properties and the orientation relative to our line of sight. AGN types such as Seyferts, quasars, and blazars exhibit different properties and at the various observed wavelengths \citep{fabian1999active}. A subset of AGN, known as Compton-thick AGN (CT AGN), is characterised by its line-of-sight obscuration, represented by a parameter called column density. The corresponding column density, $N_{\rm{H}}$ for CT AGN, is $N_{\rm{H}}\ \geq 1.5 \times 10^{24} \ \text{cm}^{-2}$ and is associated with X-ray attenuation for emission below 10 keV. This high obscuration is attributed to the presence of large amounts of dust and gas in the circumnuclear region, often related to the torus structure that can be explained by the AGN unified model \citep{antonucci1993unified}. As a result, the primary emission from AGN is almost completely enshrouded and limits our observation towards the nucleus, which leads us to rely on indirect signatures, such as scattered X-rays, IR reprocessing, and molecular gas emissions.

However, the study of CT AGN faces several challenges. Due to their obscured nature, CT AGNs are often misclassified \citep{malizia2009fraction}, especially when using the traditional tool of identifying and classifying AGN. Even when detected, the X-ray attenuation complicates the determination of the intrinsic luminosities and other properties. These limitations highlight the importance of multi-wavelength study towards CT AGN sources, which can reveal different properties of the obscuring structures around the central engine. For example, IR and sub-millimetre observations can probe the reprocessed emission from the obscuring material, while optical and radio can provide useful insights towards the AGN host and jet properties.

While significant work has been done to understand AGN, the subset of it, CT AGN, remains an elusive and under-explored type. Their extreme obscuration complicates the efforts in identifying these sources, particularly those that rely on X-ray observations. Thus, this will lead to incomplete census of CT AGN population and a limited understanding of their properties and evolutions. The obscured nature of CT AGN also means that the significant emission from the nuclear region is absorbed and re-emitted at a longer wavelength, and for this reason a multi-wavelength approach is beneficial in studying their properties more effectively. With that being said, although there are numerous studies that have utilised specific wavelengths to investigate particular aspects of CT AGN \cite[e.g.][]{akylas2012constraining, vignali2014space,lanzuisi2015compton, annuar2015nustar, sengupta2023compton, musa2025x}, a multi-wavelength approach remains less pursued.


\subsection{Multi-wavelength Approach}
A significant effort in CT AGNs is to validate the classification these sources, recognizing that CT AGNs come in different "flavours". First, the confirmed CT AGNs, for which high-quality X-ray spectra, such as those obtained in the 70-Month \textit{Swift}/BAT sample, provide direct measurements of high column density ($N_\mathrm{H} \geq 1.5 \times 10^{24} \, \rm{cm}^{-2}$). Second "flavour" consists of CT AGNs candidates, which are identified through multi-wavelength indicators, such as specific IR colours or optical line ratios, because their X-ray data are inconclusive, often due to greater distance or heavier obscuration ($N_\mathrm{H} \geq 10^{25} \, \rm{cm}^{-2}$).

The main goal of this study is to investigate the multi-wavelength properties of confirmed CT AGNs reported in the 70-Month \textit{Swift}/BAT catalogue and to provide a robust multi-wavelength template for these sources. While X-ray observation are often considered as robust method of identifying AGN, extreme column density ($N_\text{H} \geq 10^{25}$ cm$^{-2}$) can suppress even hard X-ray emission, potentially leading to significant underestimations of the AGN's true power or even complete non-detections. A multi-wavelength approach therefore serves as a complementary method to X-ray observation. AGN signatures in radio, optical, and IR, which are less sensitive to line-of-sight obscuration, would provide a more holistic view of the central engine activity. Establishing such a template is essential for evaluating the effectiveness and reliability of multi-wavelength approaches to select CT AGN when direct X-ray measurements are unavailable. This approach leads to a deeper our understanding of their nature and their role in galaxy evolution. Specifically, the obscuration in AGN may arise from two distinct processes. First, the CT obscuration may be the result of a simple orientation effect, where a standard dusty torus is viewed edge-on, blocking the central engine along the line of sight \citep{urry1995unified}. Second, it may reflect a specific evolutionary stage in which the AGN is fully enshrouded by a quasi-spherical distribution of dense gas and dust during a phase of rapid black hole growth \citep{sanders1990ultraluminous, hopkins2008cosmological}. In facilitating more complete census of these sources across cosmic history \cite[e.g.,][]{klesman2014multi, laloux2023demographics, radcliffe2021nowhere},  multi-wavelength selection techniques are also crucial for revealing key insights into the unique features of CT AGNs.

Radio observations have historically been central to AGN studies, starting with the discovery of Cygnus A \cite[e.g.][]{hargrave1974observations}, a powerful radio source associated with an accreting supermassive black hole (SMBH). Radio observation are unaffected by dust obscuration and thus provide a clean probe of jet activity and AGN-driven outflows. This allows us to study the core radio luminosity and morphology (e.g., compact core vs. extended jets/lobes) to understand the prevalence and power of jets in this heavily obscured phase. The radio luminosity threshold at 1.4 GHz \cite[typically around $10^{24} \, \mathrm{W\,Hz^{-1}}$;][]{condon1989, sadler2002radio, alban2024mapping} is commonly used to distinguish AGN from star-forming galaxies, and serves as a useful diagnostic tools for probing their energy output and environmental impact. However, the interpretation of this luminosity criterion may differ for sources in heavily obscured environments.

Infrared (IR) observations provide a crucial avenue for understanding CT AGN, as they trace the thermal emission from the circumnuclear dust (the "torus") that has reprocessed the primary AGN radiation. This allows the exploration of obscured regions where optical and X-ray signals are attenuated. For a known CT source, this IR emission acts as reliable proxy for reprocessed radiation, allowing us to quantify the intrinsic bolometric power of the hidden nucleus. Furthermore, specific techniques, such as the Wide-field Infrared Survey Explorer (WISE) colour-colour diagram \cite[e.g.,][]{assef2018wise, mateos2013uncovering, secrest2015identification, barrows2021catalog, donley2012identifying}, use these IR colours and spectral features to probe the geometry, covering factor, and composition of the obscuring dust, yielding valuable insights behind these hidden energetic processes.

Optical spectroscopy provides a unique window into the narrow line regions (NLR) and their host galaxy. Even when the central engine is hidden by thick obscuring dust and gas, the emission lines (e.g., [NII] and [OIII]) can serve as a proxy of the AGN's ionisation power. Diagnostics such as the Baldwin-Phillips-Terlevich (BPT) diagram \citep{baldwin1981classification} can be utilised to study the impact of the obscured AGNs on their hosts' gas, mapping the extent of their influence and searching for signatures of outflows or shocks.

X-ray observations, on the other hand, provide a direct and robust method for probing CT AGN, as high-energy X-rays can penetrate through the voluminous obscuring material. CT AGN are often identified through X-ray luminosities exceeding $L_{\rm{X}} \geq 10^{42} \ \rm{erg \ s^{-1}}$ \cite[e.g.,] [] {jones2024distribution, pons2014new, brightman2011xmm, hart2009x}. However, the extreme obscuration often requires careful correction for attenuation effects. These observations are critical for characterising accretion processes and estimating column densities of obscuring material, enabling distinctions between CT AGNs, less obscured AGNs, and other sources such as starburst-related X-ray emitters \citep{jackson2010revealing}.

Despite these efforts, a unified understanding of CT AGNs across all wavelengths remains lacking. While these previous studies have proven their reliability in providing useful indicators for AGN classification through single-band observations, they often prove to be insufficient when applied independently. For example, IR selections can be contaminated by star-forming regions, radio luminosities may originate from or be mixed with non-AGN activities, and even X-rays, particularly in the hard band, may miss sources with high column densities ($N_\text{H} \geq 10^{25} \, \mathrm{cm^{-2}}$).

One key challenges that remains is the absence of a fully integrated, multi-dimensional approach that combines together diagnostics from different wavelengths. Most previous studies have focused on two-dimensional comparisons such as radio versus X-ray or IR versus optical, which offer only partial views and makes it difficult to determine whether CT AGNs do truly occupy distinct regions in the broader multi-wavelength parameter space.

This study builds on the work by \citet{bar2019bat} which targeted a small sample of 28 ultra-hard X-ray-selected AGN (including choice of CT objects) to explore how multi-wavelength data can complement X-ray observations. Although their sample was limited to only a few CT AGN sources, their findings were instructive, suggesting that the diverse radio and IR properties of these objects are often linked to the effects of radiation pressure on the surrounding obscuring material. Their work serves as a key pilot study and this work fleshes it out even further, by applying a systematic, multi-dimensional analysis to a much larger and well-defined sample of CT AGN from the 70-Month \textit{Swift}/BAT catalogue to definitively characterize their unique properties.

\subsection{Our Goals}

In this work, we aim to address these gaps by conducting a systematic multi-wavelength study of CT AGNs from the 70-Month \textit{Swift}/BAT catalogue. By combining data from the radio (VLASS), IR (WISE), and optical (BASS DR2), we examine how CT AGNs behave across the entire wavelength and whether any distinctive features arise from this perspective. This work explores alternative avenue of characterising CT AGN by combining these emission properties into a single diagnostic. This allows us to investigate whether multi-wavelength diagnostics are otherwise ambiguous in a single band diagnostics.

In this work, the sections are organised as follows. Section~\ref{sec:agn_samples} describes the AGN sample used in this study, selected from the 70-Month \textit{Swift}/BAT catalogue. Section~\ref{sec:multiwavelength data} outlines the multi-wavelength data acquisition from publicly available archival datasets. We explain the sample reduction procedure in Section~\ref{sec:sample_selection}. In Section~\ref{sec:analysis_result}, we detail the methods employed in our multi-wavelength analysis and our findings. The corresponding discussions are presented in Section~\ref{sec:discussion}. Finally, a summary of our work is provided in Section~\ref{sec:conclusion}. In this paper, we adopt a flat $\Lambda$CDM cosmology with parameters $H_0 = 70 \, \text{km} \, \text{s}^{-1} \, \text{Mpc}^{-1}$, $\Omega_m = 0.3$, and $\Omega_\Lambda = 0.7$.

\section{AGN Samples}
\label{sec:agn_samples}

In this work, we utilised AGN samples from the 70-Month \textit{SWIFT}/BAT Catalogue \footnote{https://swift.gsfc.nasa.gov/results/bs70mon} \citep{ricci2017batagn} and identified CT AGN samples based on \citet{ricci2015compton}. The \textit{Swift}-BAT 70-Month Catalogue conducted a hard X-ray (14–195 keV) all-sky survey, one of the deepest and most uniform surveys of its kind. It covers 50\% of the sky down to a flux limit of $1.03 \ \times \ 10^{-11} \ \rm{erg \ s^{-1} \ cm^{-2}}$ and 90\% of the sky at $1.34 \ \times \ 10^{-11} \rm{erg \ s^{-1} cm^{-2}}$, resulting in the detection of 1170 sources, of which 838 are identified as AGN \citep{baumgartner201370}. Using this catalogue, \citet{ricci2015compton} reported 55 CT AGN sources which constitute $7.6^{+1.1}_{-2.1} \%$ of the total AGN population with average redshift of z = 0.055, excluding blazar samples (106 identified as blazars).

\section{Multi-wavelength Data}
\label{sec:multiwavelength data}

\subsection{Radio}
We obtained radio data from the Very Large Array Sky Survey \citep[VLASS;][]{lacy2020karl}. VLASS is a 2-4 GHz (S-band) survey observing the north sky ($\sim 40\degree$) with an angular resolution of $\sim$ 2.5", covering three different epochs. VLASS succeeds legacy surveys such as Faint Images of the Radio Sky at Twenty centimetres (FIRST; \citet{condon1998nrao}) and NRAO VLA Sky Survey (NVSS; \citet{becker1995first}) after 20 years. This long baseline is essential for identifying transient radio sources and understanding how AGN activity fluctuates over decades.

For this work, we used the Epoch 1 data, as presented in the catalogue from \citet{gordon2021quick}, which has a typical median RMS sensitivity of $\sim70 \, \mu \mathrm{Jy \, beam^{-1}}$. We cross-matched our \textit{Swift}/BAT sample with this catalogue to extract 2-3 GHz flux densities. For sources that were not detected, we estimated upper limits based on the local RMS noise.

\subsection{Infrared}
We obtained mid-infrared (MIR) data from the AllWISE Source Catalogue \citep{cutri2021vizier}. This catalogue provides photometry from the Wide-field Infrared Survey Explorer \citep[WISE;][]{wright2010wide}. The survey observed the entire sky in four bands: 3.4 $\mu$m (W1), 4.6 $\mu$m (W2), 12 $\mu$m (W3), and 22 $\mu$m (W4) with the angular resolution (FWHM) of $\sim$6.1" (W1), $\sim$6.4" (W2), $\sim$6.5" (W3), and $\sim$12.0" (W4) \citep{wright2010wide}.

The AllWISE catalogue, which combines data from multiple epochs, provides 5$\sigma$ point-source sensitivities of approximately 0.054, 0.071, 0.73, and 4.5 mJy in the W1, W2, W3, and W4 bands, respectively \citep{cutri2021vizier}. We cross-matched our \textit{Swift}/BAT sample with this catalogue to extract the flux densities (or upper limits) in all four WISE bands for our analysis.

\subsection{Optical}

We performed an optical analysis of our sources using archival spectroscopic data from the BAT AGN Spectroscopic Survey \citep[BASS DR2;][]{koss2023bat}, which targets local AGN ($z \lesssim 0.5$) detected in the \textit{Swift}-BAT hard X-ray survey. The second data release provides a nearly complete set of redshift and black hole mass measurements, offering a comprehensive resource for probing AGN properties across a wide range of luminosities, black hole masses, and obscuration levels \citep[e.g.,][]{koss2022bassxxi,koss2022bassxxii,koss2022bassxxvi,pfeifle2022bassxxiii,oh2022bass,mejia2022bassxxv,den2022bassxxviii,ricci2022bassxxix,ananna2022bassxxx}.

Our study uses spectroscopic measurements as reported by \citet{oh2022bass}. The latter dataset comprises 743 high-quality, unbeamed, and unlensed AGN from the \textit{Swift}-BAT 70-Month ultra-hard X-ray all-sky survey catalogue, covering 99.6\% of the non-beamed AGN sample. The survey provides optical classifications consistent with X-ray obscuration, with a dichotomy at $N_\text{H} = 10^{22}$ cm$^{-2}$, and reveals that the type 1 AGN fraction increases with intrinsic 2–10 keV luminosity.

\section{Sample Selection}
\label{sec:sample_selection}

We applied 5" and 3" cross-matching radius for VLASS and AllWISE respectively, and required complete multi-wavelength coverage. Undetected sources were adopted as upper limits and incorporated into the regression analysis using censored regression (PyMC Censored Regression model aka Tobit Model) but excluded from non-censored statistical analyses (e.g. PCA and machine learning) to maintain consistency and avoid biases from incomplete measurements. The parent sample, taken from the 70-Month \textit{Swift}/BAT catalogue, contains 838 AGNs (comprising 56 CT AGNs and 782 non-CT AGNs). Blazars were excluded due to their continuum-dominated spectra, lack of emission lines or host galaxy features, and higher redshift ($z>0.3$) broad-line quasars (see \citealt{koss2022bassxxi}).

\subsection{Uniformity of the samples}

To further refine our sample, we need to verify that our sources are drawn from the same parent population, which is crucial to deliver a uniform analysis. A bias in properties such as redshift, X-ray luminosity, BH mass ($M_{\text{BH}}$), or Eddington Ratio ($\lambda_{\text{Edd}}$) could potentially introduce discrepancies in multi-wavelength comparison analyses, especially for parameters that involve absolute luminosities.

\begin{figure*}
    \centering
    \includegraphics[width=1\linewidth]{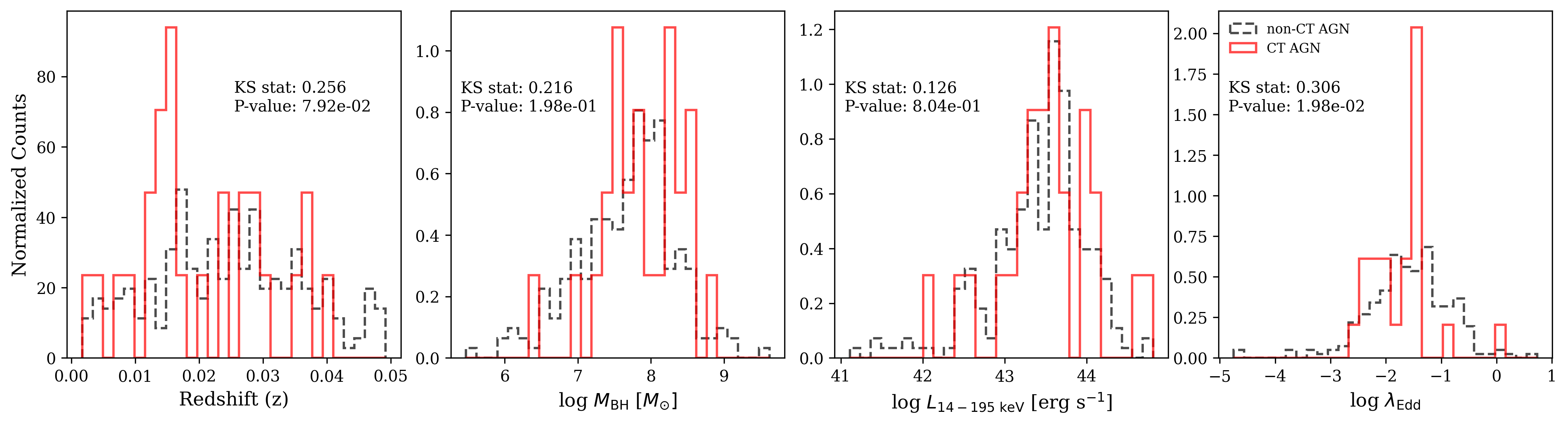}
    \caption{Comparison of the normalized distributions of redshift (z), black hole mass (log $M_\text{BH}$), X-ray luminosity (log $L_{14-195}$ keV), and Eddington ratio (log $\lambda_\text{Edd}$) for the CT AGN (red solid lines) and non-CT AGN (black dashed lines) samples.}
    \label{fig:uniform_dist}
\end{figure*}

We applied a redshift limit to $z\leq 0.05$ which corresponds to the average redshift of CT AGN sample from \citet{ricci2015compton}. We further verify this by comparing the distribution of the properties: redshift ($z$), BH mass ($M_\text{BH}$), X-ray luminosity ($L_\mathrm{14-195\,keV}$) and Eddington ratio ($\lambda_\mathrm{Edd}$). The normalized distribution for the samples are shown in Figure~\ref{fig:uniform_dist}

The K-S test for the redshift distribution reported a $p$-value of 0.079 confirming that both CT and non-CT AGN were drawn from the same population. Similarly, we found no statistical differences in BH mass ($p$ = 0.216) and X-ray luminosity ($p$ = 0.126). This confirms that our sample are distributed within the same population and well-matched distance scale.

However, we found a significant difference in the Eddington ratio ($p$ = 0.02), where CT AGNs show slightly higher accretion rates on average. Since the black hole masses are consistent, this likely reflects a real physical difference rather than a sample bias. This fits with the evolutionary picture where high obscuration is linked to phases of rapid growth and higher Eddington ratios \citep[e.g.,][]{fabian1999active, ricci2017batagn}. We therefore proceed with the analysis while noting this intrinsic distinction.

This results in a final sample of 243 sources (26 CT AGN, 217 non-CT AGN), with the available data are: (i) radio flux – VLASS (2–3 GHz); (ii) IR flux and luminosities - (WISE W1–W4); (iii) optical emission lines - H$\alpha$, H$\beta$, [OIII]$\lambda5007$, [SII]$\lambda\lambda6716+6730$, [NII]$\lambda6584$, [OI]$\lambda6300$; (iv) X-ray - $L_\text{int, 2–10 keV}$, $L_\text{int, 20-50 keV}$, $L_\text{int, 14-195 keV}$, $N_{\rm H}$, and Eddington ratio ($\lambda_\text{Edd}$). This uniform dataset enables consistent and directly comparable multi-wavelength analyses.

\section{Data Analysis and Results}
\label{sec:analysis_result}

\subsection{Radio Properties}
\label{subsec:radioproperties}

For the radio analysis, we utilised data from the VLASS 2-3 GHz survey, acquiring 243 sources from the catalogue. We extracted the radio flux and derived the radio luminosities ($L_{2-3\ \rm{GHz}}$) using the standard luminosity conversion by adopting a spectral index of $\alpha = -0.71$ for VLASS \citep{gordon2021quick}. The necessary redshift measurements were taken from the 70-Month Swift/BAT catalogue \cite[see][Table 1]{ricci2017batagn}, and a $K$-correction was applied to account for the redshift dependence.

\subsubsection{Radio Luminosities}
\label{subsubsec:Lradio}
\begin{figure}
    \centering
    \includegraphics[width=\linewidth]{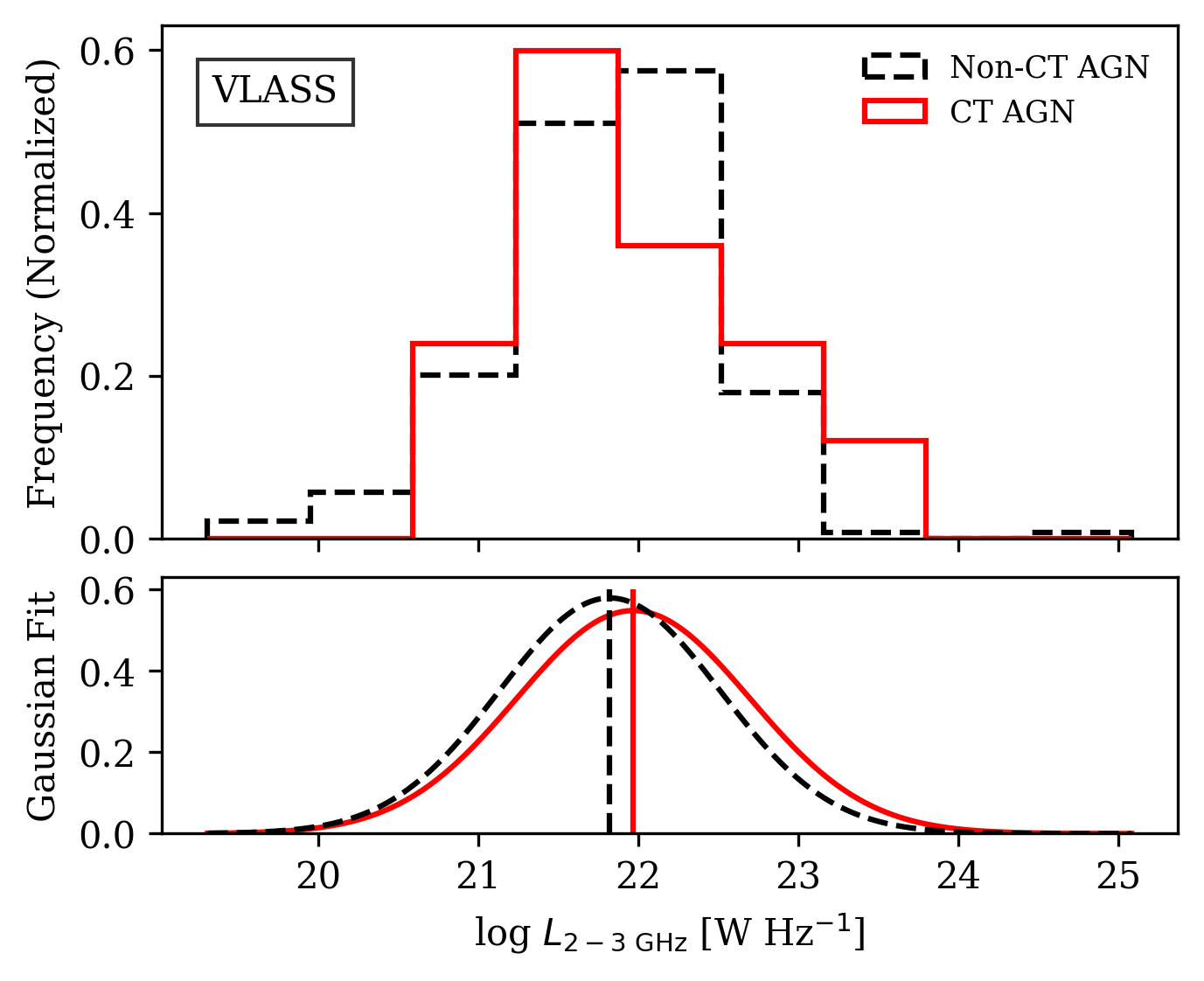}
    \caption{Distribution of  $L_{2-3\ \rm{GHz}}$ for different AGN classes. The histogram on the top panel shows the luminosity distribution for non-CT AGN (black dashed line) and CT AGN (red solid line). The bottom panel show the fitted Gaussian for the AGN population with the vertical lines represent the fitted mean radio luminosity, $\mu$.}
    \label{fig:Lradio_histogram}
\end{figure}

To explore the distribution characteristics of our sample, we applied a Gaussian fitting model to the radio luminosity histogram, as shown in Figure \ref{fig:Lradio_histogram}. The fitted parameters, including the mean and standard deviation, are summarised in Table \ref{tab:fitted params}. 

At 2-3 GHz, the luminosity distributions of CT and non-CT AGN are broadly similar, with both peaking near 10$^{22}$ W Hz$^{-1}$ and spanning approximately 10$^{20}$ - 10$^{25}$ W Hz$^{-1}$. The Gaussian fits indicate that CT AGN exhibit a slightly higher mean radio luminosity compared to non-CT AGN sample, although they maintain a relatively narrow distribution. The difference in these mean values is not statistically significant within the uncertainties (Table~\ref{tab:fitted params}). Since VLASS angular resolution is approximately 2.5", the observed emission is primarily attributed to the compact central regions.

\begin{table}
    \centering
    \caption{Mean ($\mu$) and standard deviation ($\sigma$) of the radio luminosities $L_\text{2-3 GHz}$ for the CT and non-CT AGN. The $\mu$ values are presented with their uncertainties, while $\sigma$ represents the standard deviation of the distributions.}
    \renewcommand{\arraystretch}{1.1}
    \begin{tabular}{lcc}
        \hline
        \multirow{3}{*}{Sample}   & \multicolumn{2}{c}{$L_{2-3 \ \rm{GHz}}$} \\ [2pt]
        \cline{2-3}
        & $\mu$ & $\sigma$ \\ [2pt]
        & (log W Hz$^{-1}$) & (log W Hz$^{-1}$) \\
        \hline
        Non-CT AGN & 21.84 $\pm$ 0.05& 0.69\\
        CT AGN     & 21.97 $\pm$ 0.14& 0.73\\
        \hline
    \end{tabular}
\label{tab:fitted params}
\end{table}

\subsubsection{Radio Luminosities versus X-ray Luminosities}
\label{subsubsec:LradiovsLxray}

\begin{figure*}
    \centering
    \includegraphics[width=\linewidth]{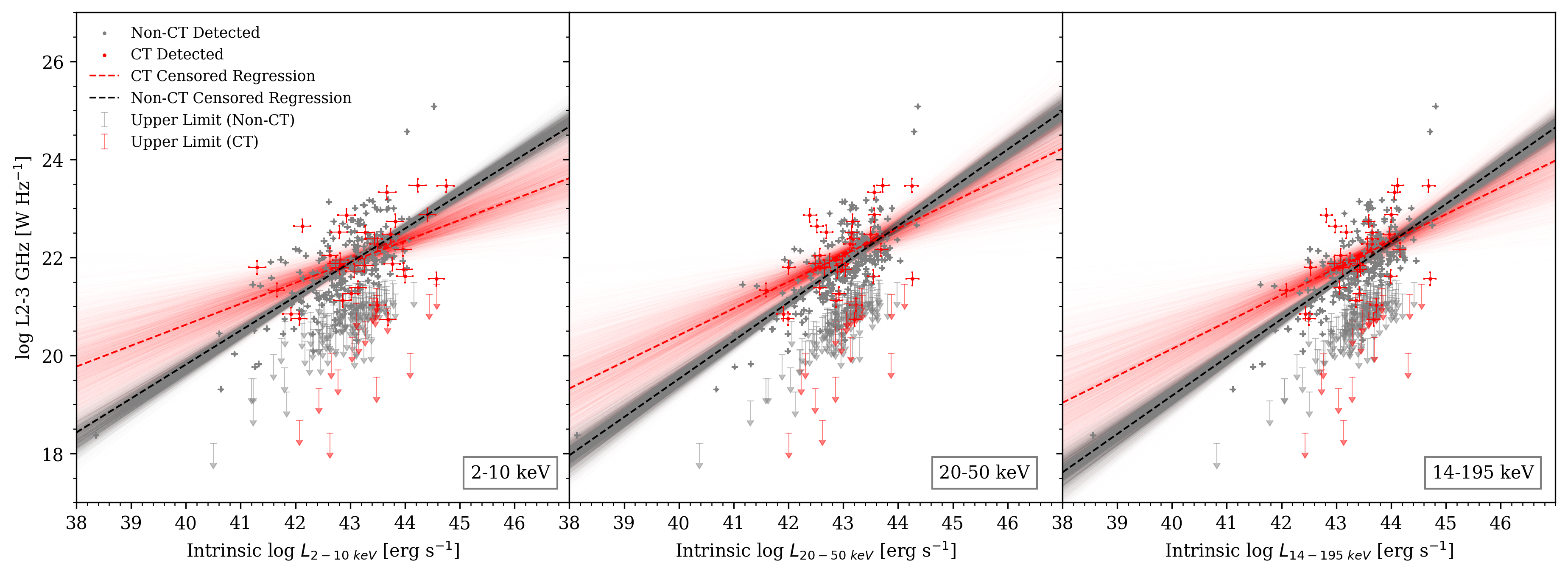}
    \caption{Correlation between intrinsic radio luminosity, log $L_{\text{2–3 GHz}}$, and X-ray luminosities across three energy bands: 2–10 keV (left), 20–50 keV (middle), and 14–195 keV (right). Non-CT AGN are shown in grey and CT AGN in red, with downward arrows indicating upper limits for non-detections. The dashed lines represent the censored regression fits, with grey for non-CT AGN and red for CT AGN. The faint lines around each regression indicate posterior draws, highlighting the uncertainty in the fitted relations.
}
    \label{fig:lxray-lradio}
\end{figure*}

To further look into how radio luminosities vary throughout X-ray bands, we explored its correlation with different X-ray luminosity bands. The correlation study was performed using statistical tests such as Kendall's ($\tau$) and Spearman's ($\rho$) on the calculated radio luminosities and the X-ray luminosities. The X-ray luminosities were taken from the same catalogue from our sample, the 70-Month \textit{Swift}/BAT Catalogue \cite[see][Table 13]{ricci2017batagn}. The X-ray and radio luminosities correlation is presented in Figure~\ref{fig:lxray-lradio}. Both parameters are displayed in logarithmic scale. In the regression analysis, we explicitly account for upper limits in radio luminosities by applying a censored regression framework (Tobit regression) implemented in \texttt{PyMC}, ensuring that non-detections contribute information to the fitted relations.

\begin{table*}
    \centering
    \caption{Spearman ($\rho$) and Kendall ($\tau$) correlation coefficients for the X-ray luminosities, $L_X$ 2-10 keV, 20-50 keV, and 14-195 keV for CT and non-CT AGN in the VLASS samples. The values in parentheses indicate the corresponding p-values.}
    \setlength{\tabcolsep}{4pt}
    \renewcommand{\arraystretch}{1.2}
    \begin{tabular}[c]{lcccccc}
        \hline
        \multirow{2}{*}{Sample} & \multicolumn{2}{|c|}{$L_{\text{X}\sim2-10\, \text{keV}}$} & \multicolumn{2}{|c|}{$L_{\text{X}\sim20-50\, \text{keV}}$} & \multicolumn{2}{|c|}{$L_{\text{X}\sim14-195\, \text{keV}}$} \\
        \cline{2-7}
        & $\rho$ & $\tau$ & $\rho$ & $\tau$ & $\rho$ & $\tau$ \\
        \hline
        Non-CT AGN & 0.51 ($1.82\times10^{-26}$) & 0.36 ($2.74\times10^{-25}$) & 0.55 ($3.79\times10^{-32}$) & 0.39 ($3.88\times10^{-30}$) & 0.56 ($5.32\times10^{-33}$) & 0.40 ($1.07\times10^{-30}$)\\
        CT AGN & 0.30 ($3.16\times10^{-2}$) & 0.20 ($4.29\times10^{-2}$) & 0.30 ($3.22\times10^{-2}$) & 0.22 ($2.61\times10^{-2}$) & 0.28 ($5.20\times10^{-2}$) & 0.20 ($4.29\times10^{-2}$)\\
        \hline
    \end{tabular}
\label{tab:vlasslx_lr}
\end{table*}

Across all X-ray bands, both CT and non-CT AGN display positive correlations with radio luminosities (Figure~\ref{fig:lxray-lradio}) consistent with previous findings (\citealp[e.g.][]{vink2006x, panessa2007x, li2008black, bianchi2009caixa, pasini2020relation}). For non-CT AGNs, the Spearman ($\rho$) and Kendall ($\tau$) tests (Table~\ref{tab:vlasslx_lr}) reveal good and significant correlations $\rho \approx$  0.51 - 0.56, $\tau \approx$ 0.36 - 0.40) with $p$-values of $10^{33}-10^{25}$, confirming a robust link between radio and X-ray emissions. In contrast, CT AGNs show only moderate correlations \textcolor{red}{($\rho \approx$ 0.28 – 0.30, $\tau \approx$ 0.20 – 0.22)} with $p$-values of order $10^{-2}$. The visual separation between non-CT and CT is evident across all three X-ray bands while the regression fits display comparable scatter and uncertainties.

However, the K-S test on the distribution of the X-ray-to-radio-luminosity ratio shows that the population is not statistically different across the X-ray bands. The K-S test value for $L_\text{2-10\,keV}$ ($p=0.157$), $L_\text{20-50\,keV}$ ($p=0.239$), and $L_\text{14-195\,keV}$ ($p=0.261$) fail to reject the null hypothesis, showing no significance difference between the distributions. The regression fits across all X-ray bands display comparable scatter and uncertainties.

\subsection{WISE Luminosities}

\begin{figure*}
    \centering
    \includegraphics[width=1\linewidth]{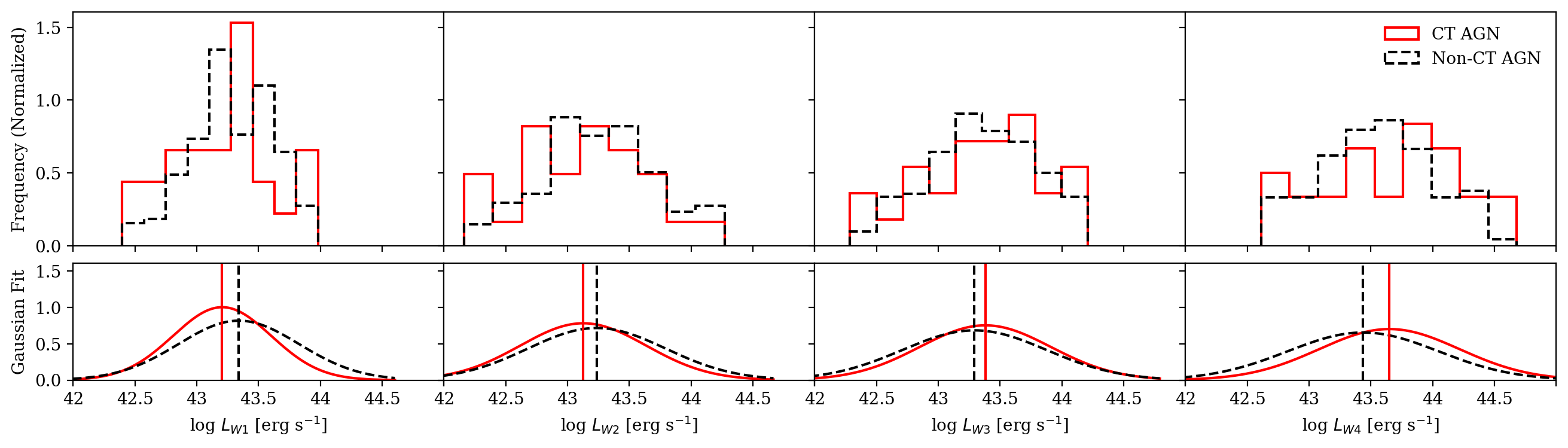}
    \caption{WISE band luminosity distributions across different AGN populations. Non-CT AGNs are shown in blue, CT AGNs in red, and blazars in green. The top panel shows the histogram of the luminosity distributions, and the fitted Gaussian with the mean luminosity, $\mu$, are shown in the bottom panel.}
    \label{fig:wise_luminosities}
\end{figure*}

To study the intrinsic IR luminosities throughout our AGN samples, we derived the WISE luminosities across all four bands (W1, W2, W3, and W4). We obtained the WISE flux by adopting the zero-point flux magnitude from \citet{jarrett2011spitzer}, assuming a power-law spectrum ($f_v \propto v^\alpha$) with a spectral index ($\alpha=-1$; \citealt{wright2010wide}), and applying a K-correction to account for the redshift dependence. The resulting luminosity values are presented in units of $\text{erg s}^{-1}$. Figure~\ref{fig:wise_luminosities} shows the luminosity distribution of the WISE bands for our sample, with the fitted Gaussian parameters listed in Table \ref{tab:wise_luminosities}.

The fitted distribution shows substantial overlap between both populations, indicated by mean luminosities with no significant offset between them. In the short wavelengths, W1 and W2, the mean luminosities of the CT AGNs slightly fall below the non-CT mean luminosities. Conversely, towards longer wavelengths, W3 and W4, the CT AGNs distribution is slightly shifted to higher luminosities. Throughout all the bands, the fitted mean luminosities remain comparable between CT and non-CT AGNs within the uncertainties.

\begin{table*}
    \centering
    \caption{Mean luminosities ($\mu$) and standard deviation ($\sigma$), of the WISE W1, W2, W3, and W4 bands for both CT and non-CT AGNs. The $\mu$, are presented with their uncertainties, and $\sigma$ represents the standard deviation of the luminosity distributions. All luminosities are in units of erg\,s$^{-1}$.}
    \renewcommand{\arraystretch}{1.1}
    \begin{tabular}{lcccccccc}
        \hline
        \multirow{3}{*}{Sample} & \multicolumn{2}{c}{W1} & \multicolumn{2}{c}{W2} & \multicolumn{2}{c}{W3} & \multicolumn{2}{c}{W4} \\
        \cline{2-9}
         & $\mu$ & $\sigma$ & $\mu$ & $\sigma$ & $\mu$ & $\sigma$ & $\mu$ & $\sigma$ \\ [1pt]
         & (log erg\,s$^{-1}$) & (log erg\,s$^{-1}$) & (log erg\,s$^{-1}$) & (log erg\,s$^{-1}$) & (log erg\,s$^{-1}$) & (log erg\,s$^{-1}$) & (log erg\,s$^{-1}$) & (log erg\,s$^{-1}$) \\
        \hline
        Non-CT AGN & 43.67 $\pm$ 0.03 & 0.64 & 43.60 $\pm$ 0.04 & 0.71 & 43.63 $\pm$ 0.04 & 0.70 & 43.77 $\pm$ 0.04 & 0.71 \\
        CT AGN     & 43.21 $\pm$ 0.08 & 0.40 & 43.13 $\pm$ 0.10 & 0.51 & 43.38 $\pm$ 0.10 & 0.53 & 43.65 $\pm$ 0.11 & 0.57 \\
        \hline
    \end{tabular}
\label{tab:wise_luminosities}
\end{table*}

\subsection{WISE Colour Distributions}
\begin{figure}
    \centering
    \includegraphics[width=1\linewidth]{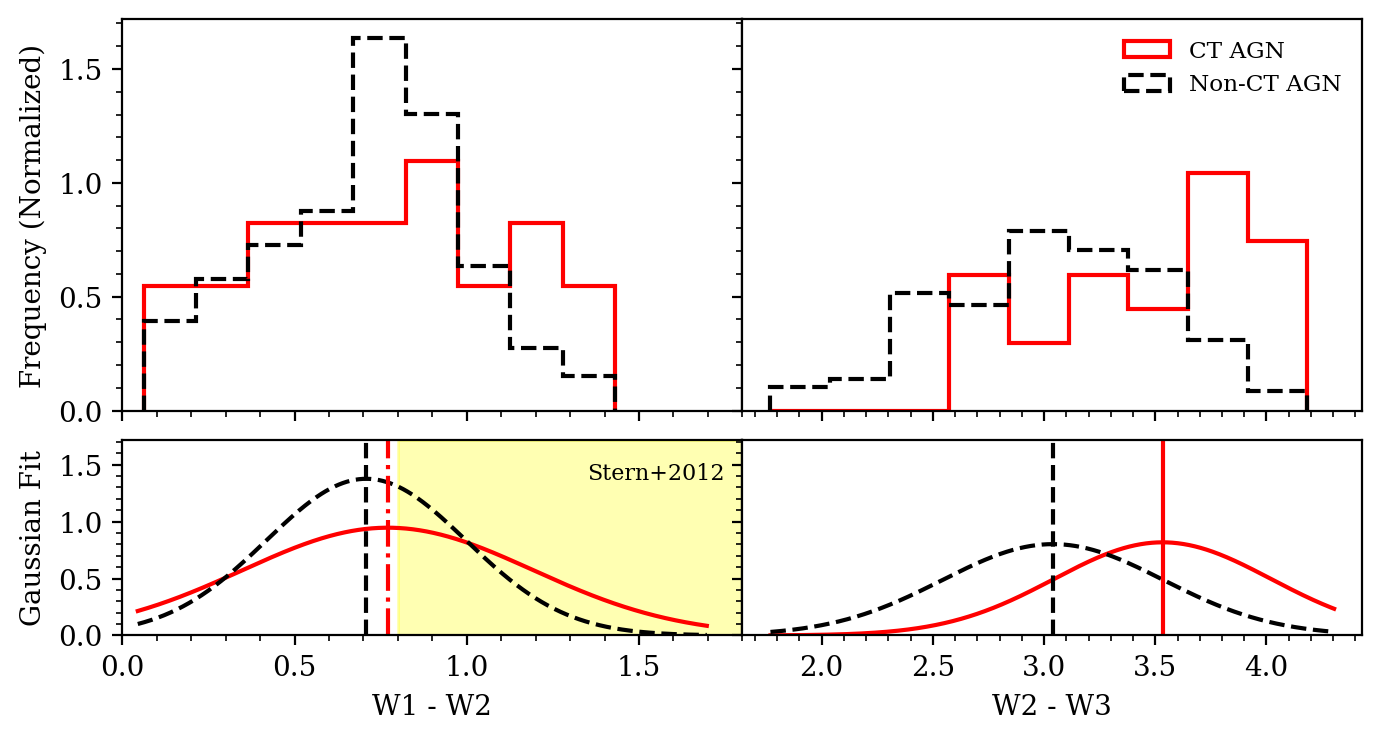}
    \caption{W1-W2 (left panel) and W2-W3 (right panel) colour diagrams for different AGN samples. The W1-W2 colour cut AGN selection criteria from \citet{stern2012mid} is shown in the yellow region.}
    \label{fig:w1w2_w2w3}
    
\end{figure}

To characterize the mid-IR properties, we examined the $W1-W2$ and $W2-W3$ colour distributions, which serve as proxies for the hot-dust and cool-dust contributions, respectively. For AGN selection, the $W1-W2 \geq 0.8$ threshold is the most widely used selection criterion. This threshold is effective at minimising contamination from host galaxies, particularly in low-luminosity sources \citep{stern2012mid, daoutis2023versatile}. We applied a Gaussian fitting to the distributions shown in Figure \ref{fig:w1w2_w2w3} to quantify the mean value ($\mu$) and standard deviation ($\sigma$). The fitted parameters are summarised in Table \ref{tab:w1w2_w3w4_hist}.

\subsubsection{W1--W2 Colour}
The $W1-W2$ colour distribution illustrates the AGN selection properties relative to the \citet{stern2012mid} criterion ($W1-W2 \geq 0.8$). Both CT and non-CT AGNs display overlapping distributions with comparable mean values of $\mu = 0.77$ and $\mu = 0.80$, respectively. However, the populations differ in their spread: the CT AGNs exhibit a broader dispersion ($\sigma = 0.42$) compared to the non-CT AGNs ($\sigma = 0.29$).

\subsubsection{W2--W3 Colour}
In contrast to the shorter wavelengths, the $W2-W3$ colour distribution reveals a noticeable shift. CT AGNs tend to exhibit redder colours compared to non-CT AGNs, indicating a stronger relative contribution from the longer-wavelength W3 band in the obscured population.

\begin{table}
    \centering
    \caption{Mean ($\mu$) and standard deviation ($\sigma$) of the WISE $W1-W2$ and $W2-W3$ colours for CT AGNs and non-CT AGNs. The mean values are presented with their associated uncertainties, while $\sigma$ represents the standard deviation of the distributions.}
    \renewcommand{\arraystretch}{1.1}
    \begin{tabular}{lcccc}
        \hline
        \multirow{3}{*}{Sample} & \multicolumn{2}{c}{$W1-W2$} & \multicolumn{2}{c}{$W2-W3$} \\
        \cline{2-5}
        & $\mu$ & $\sigma$ & $\mu$ & $\sigma$ \\ [2pt]
        & (mag) & (mag) & (mag) & (mag) \\
        \hline
        Non-CT AGN & 0.71 $\pm$ 0.02 & 0.29 & 3.04 $\pm$ 0.03 & 0.50 \\
        CT AGN     & 0.77 $\pm$ 0.08 & 0.42 & 3.53 $\pm$ 0.10 & 0.49 \\
        \hline
    \end{tabular}
    \label{tab:w1w2_w3w4_hist}
\end{table}

\subsubsection{WISE Colour-colour Diagram}
\begin{figure*}
    \centering
    \includegraphics[width=0.9\linewidth]{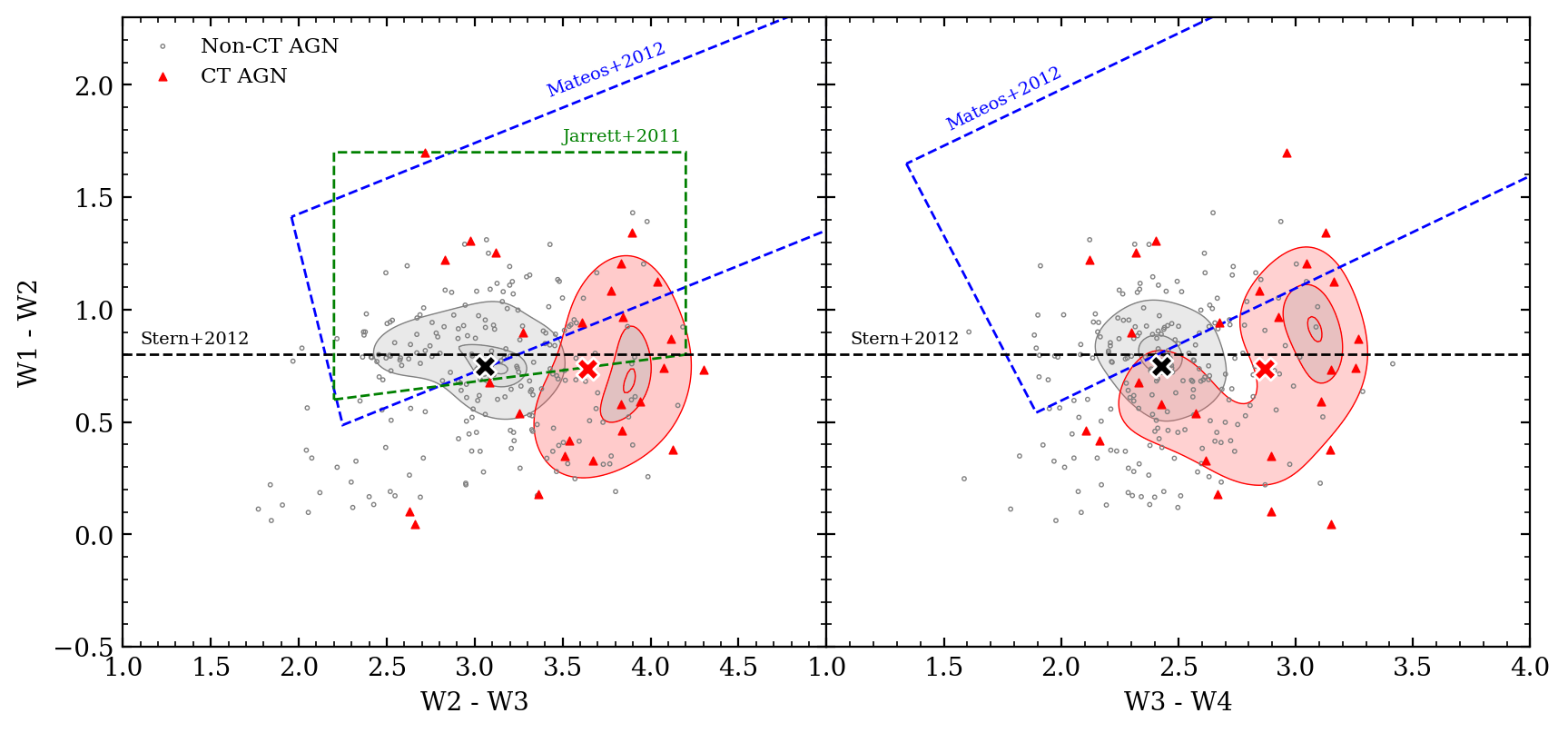}
    \caption{WISE colour-colour diagrams using three (left panel) and four (right panel) WISE bands for our sample, comprising of CT AGNs (red triangles) and non-CT AGNs (grey circles). The AGN selection wedges/criteria from \citet{stern2012mid} (black), \citet{jarrett2011spitzer} (green), and \citet{mateos2012using} (blue) are indicated by the dashed lines. The contours represent the 1$\sigma$, 2$\sigma$, and 3$\sigma$ of the fiducial density, with the barycentre of each population is marked with crosses (CT in red and non-CT in black).}
    \label{fig:wise_diagram}
\end{figure*}

We investigate the IR properties of our AGN samples using the standard WISE AGN selection diagnostic, incorporating all WISE bands (W1, W2, W3, and W4) and adopt AGN selection criteria from the past literature (\citealp[e.g.][]{jarrett2011spitzer, stern2012mid, mateos2013uncovering}). The WISE is useful to distinguish stars, star-forming galaxies, normal galaxies, and AGNs. In this work, we plotted two types of WISE diagrams, which use 3 WISE bands and 4 WISE bands, shown in Figure~\ref{fig:wise_diagram}. This approach follows the previous work by \citet{mateos2012using} to investigate the completeness of AGN selection techniques using both types of WISE diagrams. In addition to the W4 band, it displays shallower depth compared to the first three bands and suffers contamination from star-forming galaxies. Despite this restriction, we wanted to explore the CT AGNs distribution in this AGN selection as compared to the non-CTs.

To verify whether this distribution is physically real or an artifact of the small sample size, we performed a bootstrap scatter analysis (Figure~\ref{fig:wise_bootstrap}). We compared the dispersion of our 26 CT sources against 10,000 random subsets of 26 non-CT sources. The analysis confirms that the increased scatter in CT AGNs is statistically significant. For the 3-band diagram, we obtained a $p$-value of 0.0441. For the 4-band diagram, the difference is highly significant with $p = 0.0001$.

The fiducial density of both population is represented by the contours, showing the 68\%, 97\% and 99\% distribution levels (1$\sigma$, 2$\sigma$, and 3$\sigma$ of the distribution) with the barycentre marked as a large cross. In both diagrams, the CT and non-CT AGNs exhibit a broadly overlapping distribution. We find that the AGN selection wedge proposed by \citep{mateos2012using} captures a substantial fraction of both CT and non-CT AGNs. However, we observed a subset of CT AGNs that scatters outside the primary selection region, particularly extending towards lower $W1-W2$ colours and slightly elevated $W2-W3$ or $W3-W4$ colours. The barycentre of CT AGNs showed offset position from non-CT, shifting towards the redder colours in the $W2-W3$ and $W3-W4$.

\begin{figure}
    \centering
    \includegraphics[width=\linewidth]{figures/wise_bootstrap.png}
    \caption{Bootstrap scatter analysis in the WISE colour-colour diagram for our sample. The histogram illustrates the mean scatter distance from the centroid, comparing CT AGNs (red dashed line) with non-CT AGNs (black dash-dotted line). The upper panel shows the three-band WISE diagram distribution, while the bottom panel shows the four-band WISE diagram.}
    \label{fig:wise_bootstrap}
\end{figure}

\subsection{Optical Emission Line Properties}
We investigated the optical properties of our sample using archival spectral line measurements from the BAT AGN Spectroscopic Survey (BASS) Data Release 2 \citep{koss2023bat, oh2022bass}. We obtained 243 cross-matched sources with available measurements for key emission lines: [NII]$\lambda$6584, [SII]$\lambda\lambda$6716+6730, [OIII]$\lambda$5007, [OI]$\lambda$6300, H$\alpha$, and H$\beta$.

\subsubsection{Baldwin-Philips-Terlevich Diagram}

We classified the sources using three standard Baldwin-Phillips-Terlevich (BPT) diagrams: (1) [NII]$\lambda$6583/H$\alpha$ versus [OIII]$\lambda$5007/H$\beta$ (hereafter BPT-NII; \citealt{baldwin1981classification}), (2) [SII]$\lambda\lambda$6716,6730/H$\alpha$ versus [OIII]$\lambda$5007/H$\beta$ (VO87-SII) (hereafter BPT-SII; \citealt{veilleux1987spectral}), and (3) [OI]$\lambda$6300/H$\alpha$ versus [OIII]$\lambda$5007/H$\beta$ (hereafter BPT-OI; \citealt{veilleux1987spectral}). In all three diagnostics, the majority of both CT and non-CT AGNs reside in the Seyfert region. A subset of both populations falls into the LINER or star-forming/composite regions. Optical spectroscopy is sensitive to obscured AGN where the dusty medium is torus-shaped (toroidally distributed, with a certain opening angle), but it is insensitive to "buried" AGN surrounded by dust in virtually all directions \cite[e.g.][]{imanishi2006infrared, imanishi2008systematic}.

The barycentre for each BPT diagram is calculated and marked in red crosses for CT and black crosses for non-CT, with both overlapping significantly within the Seyfert region across all three diagrams. We further included the fiducial density contour at 1$\sigma$, 2$\sigma$, and 3$\sigma$ of the distribution levels. These contours reveal that the core distributions at 1$\sigma$ (68\%) of both populations are showing nearly identical shape and location.

\begin{figure}
    \centering
    \includegraphics[width=1\linewidth]{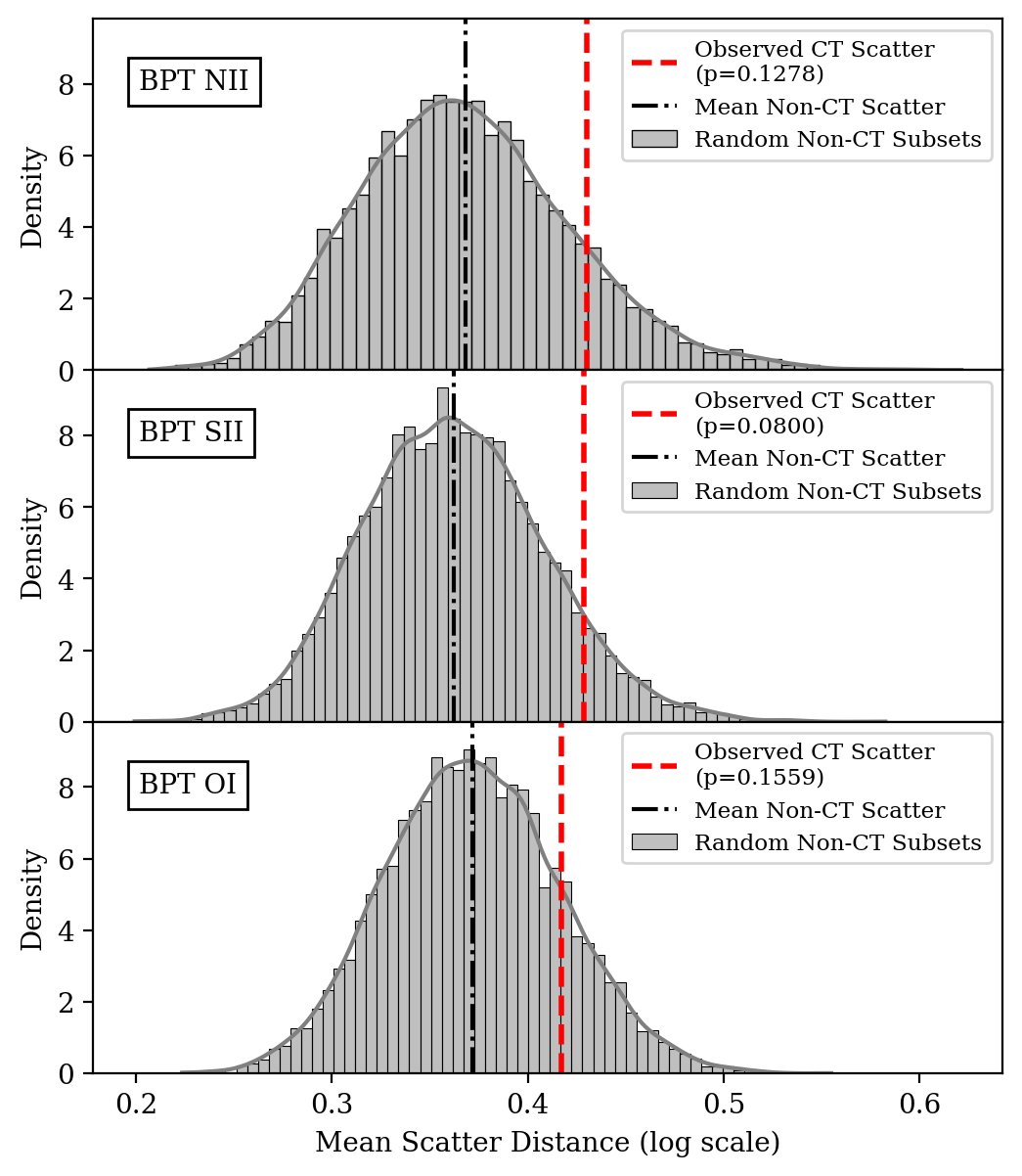}
    \caption{Bootstrap scatter analysis for all three BPT diagrams. The grey histogram illustrates the mean scatter distribution from the centroid, comparing non-CT AGNs (black dash-dotted line) with CT AGNs (red dashed line)}
    \label{fig:bpt_bootstrap}
\end{figure}

The CT AGNs ostensibly exhibit a likely larger scatter in the BPT diagrams compared to the underlying non-CT population. To verify if this scatter signifies distinct physical conditions, we performed a bootstrap resampling analysis. Similar to the IR analysis, we randomly extracted 26 sources from the non-CT sample 10,000 times and calculated the dispersion for each iteration (Figure~\ref{fig:bpt_bootstrap}). The results showed $p$-values of 0.123 (BPT-NII), 0.083 (BPT-SII), and 0.158 (BPT-OI), indicating no statistically significant difference in dispersion between the CT and non-CT samples.

\begin{figure*}
    \centering
    \includegraphics[width=1\linewidth]{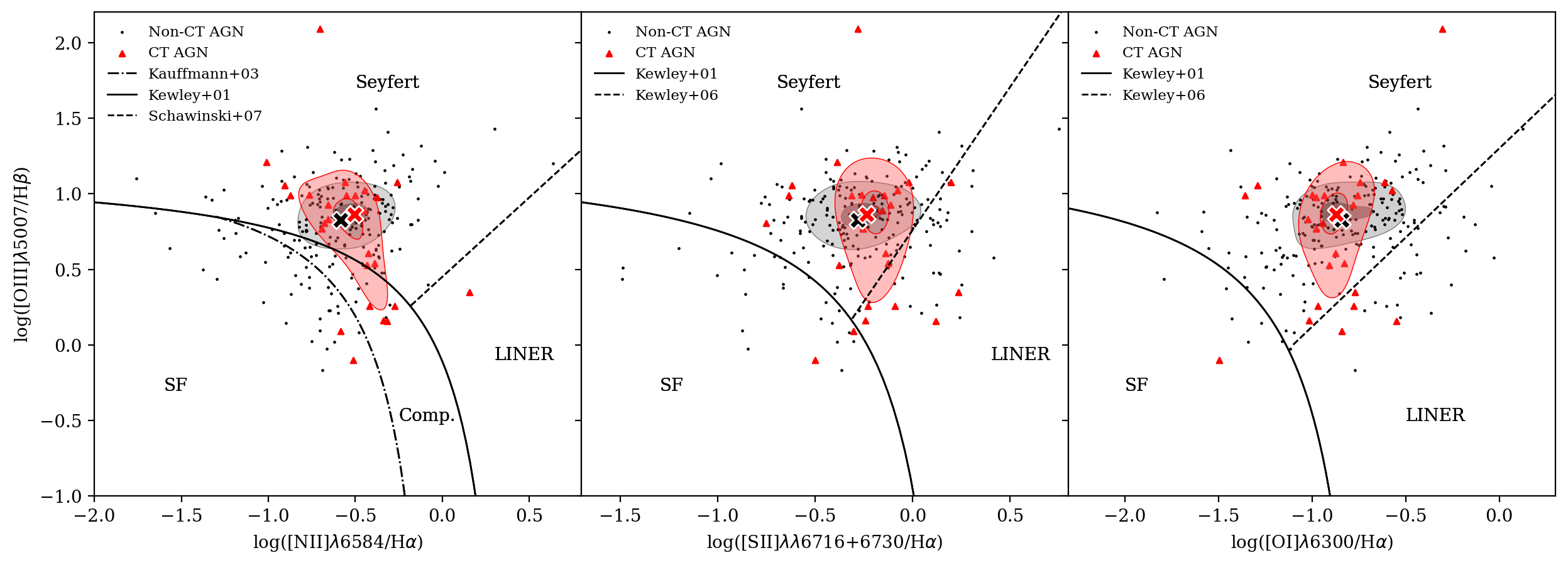}
    \caption{BPT diagrams for our AGN sample based on emission line ratios. The left panel shows the BPT-NII diagram, while the middle and right panels show the BPT-SII and BPT-OI, respectively. Demarcation lines from \citet{kewley2001theoretical} (solid), \citet{kauffmann2003host}(dashed-dotted), and \citet{schawinski2007observational}(dashed) are adopted for BPT-NII, while in BPT-SII and BPT-OI, \citet{kewley2006host}(dashed) and \citet{kewley2001theoretical}(solid) are adopted. The demarcation lines are used to distinguish AGN from star-forming regions, composite regions and LINERs. CT AGNs are shown in a red triangle, non-CT AGNs in a grey circle, and blazars in a green square. The contours represent the fiducial density, with 1$\sigma$, 2$\sigma$, and 3$\sigma$ of the population. The barycentres are marked with cross for both CT (red) and non-CT AGNs (black).}
    \label{fig:bpt_diagram}
\end{figure*}

\subsubsection{[NII]/H\texorpdfstring{$\alpha$}{alpha} versus Eddington ratio, \texorpdfstring{$\lambda_{Edd}$}{lambdaEdd}}

\begin{figure}
    \centering
    \includegraphics[width=1\linewidth]{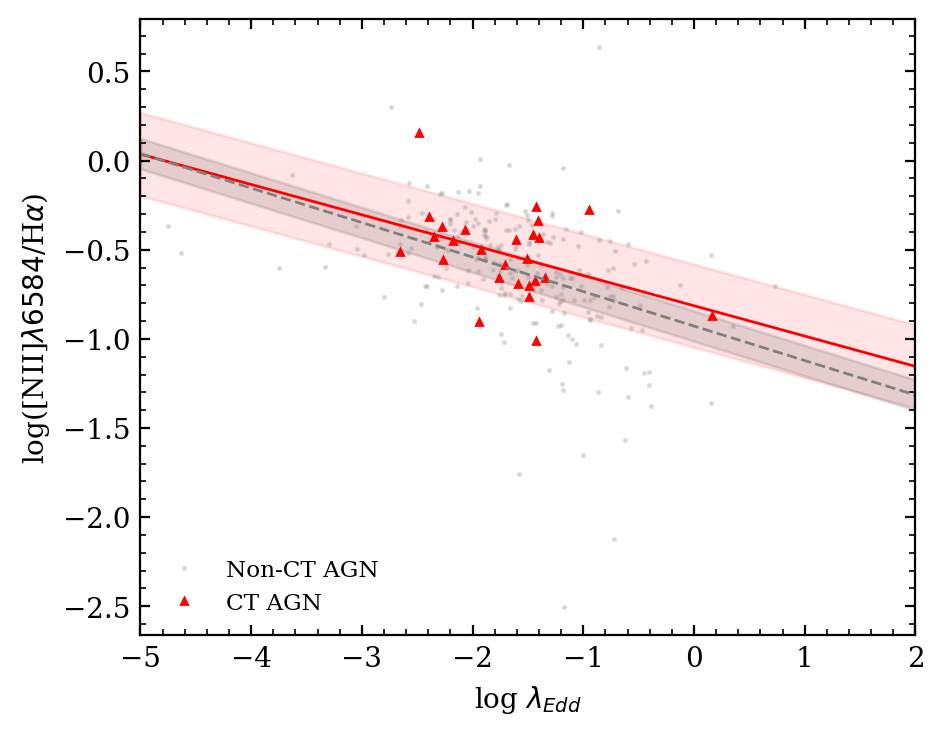}
    \caption{The relationship between Eddington ratio ($\lambda_{\rm Edd}$) and [NII]$\lambda6584$/H$\alpha$ emission-line ratio is displayed in log scale for CT AGNs (red triangles) and non-CT AGNs (grey dots). The solid red line denotes the best-fit linear regression for CT AGNs, with a shaded region indicating the $3\sigma$ uncertainty. The dashed grey line represent regression fits for non-CT AGNs.}
    \label{fig:nii_edd}
\end{figure}

We further explored the relationship between the [NII]$\lambda6584$/H$\alpha$ ratio and the Eddington ratio ($\lambda_\text{Edd}$), as shown in Figure \ref{fig:nii_edd}. Both populations exhibit a negative correlation. For non-CT AGNs, the correlation is strong (Spearman $\rho = -0.53$), while for CT AGNs, it is moderate ($\rho = -0.23$).

\subsection{Principal Component Analysis}
\label{sec:pca_analysis}

\begin{figure}
    \centering
    \includegraphics[width=1\linewidth]{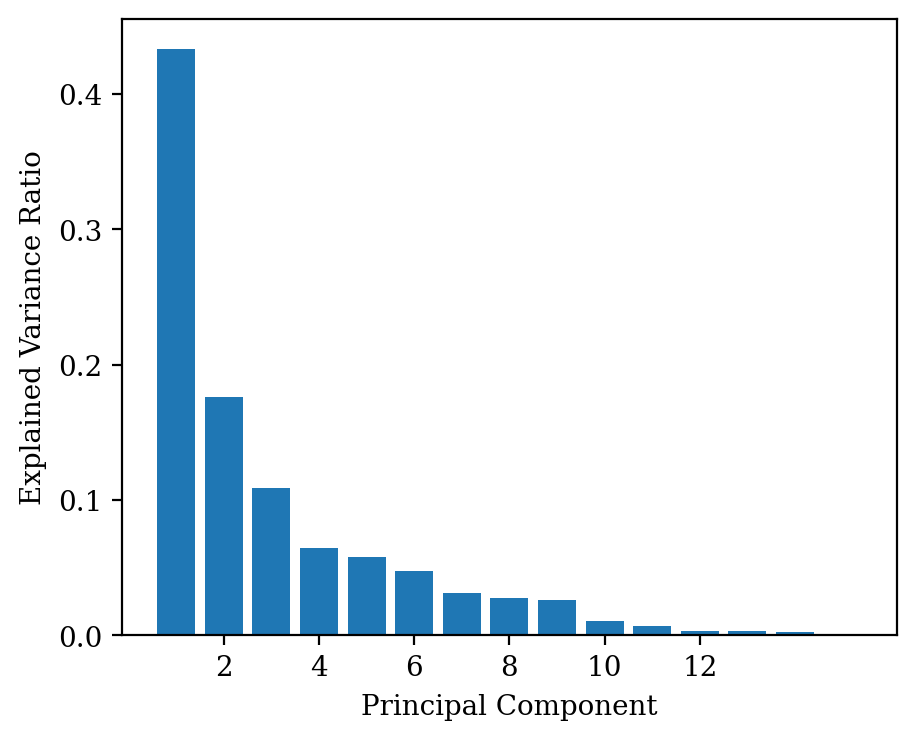}
    \caption{Scree plot of the explained variance ratio for each principal component. The first three components account for the majority of the variance in the dataset, capturing more almost 72\% of the total variance. An elbow trend is observed around the third to fourth component, suggesting that these components retain the most informative features of the data.}
    \label{fig:scree_plot}
\end{figure}

\begin{figure*}
    \centering
    \includegraphics[width=0.85\linewidth]{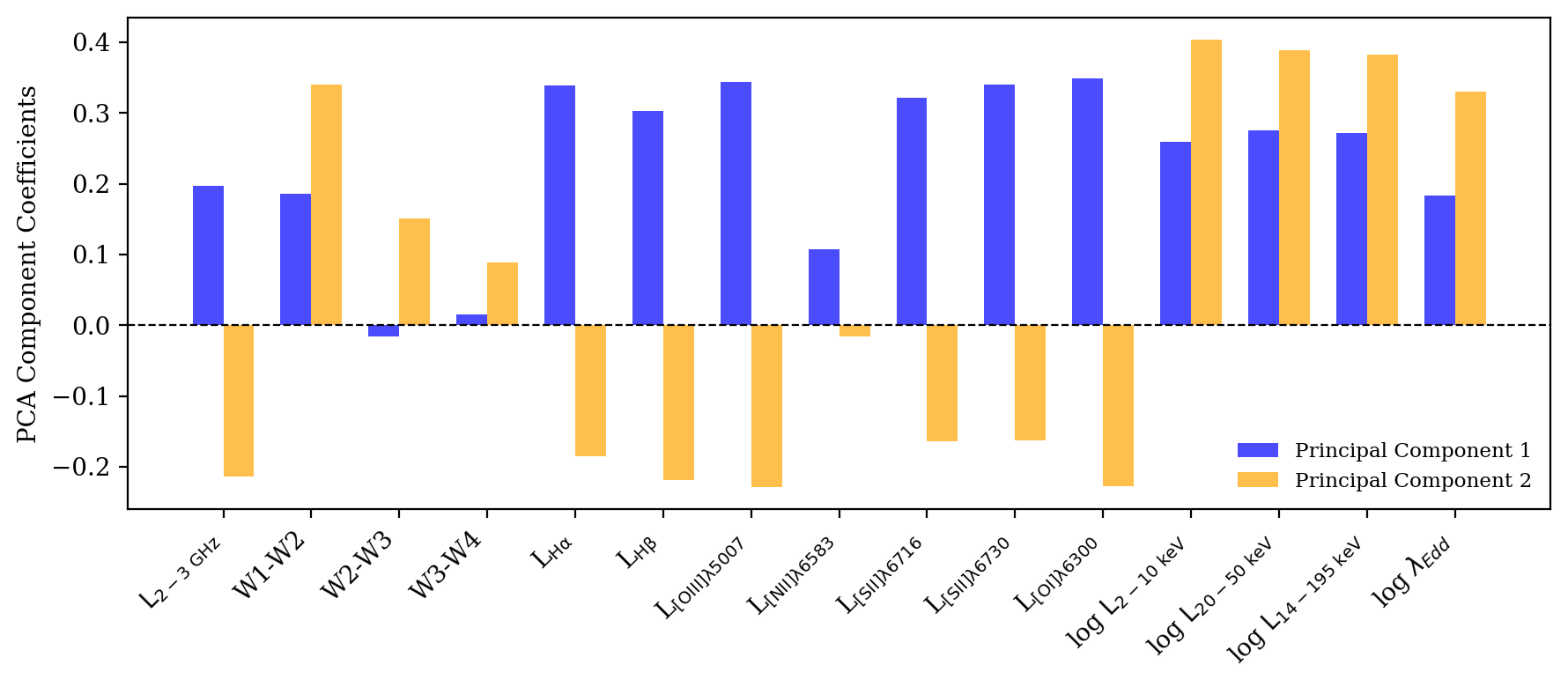}
    \caption{Principal component coefficients for the multi-wavelength features of the AGN sample. Principal Component 1 (blue) is dominated by optical emission-line ratios and X-ray luminosity. Principal Component 2 (orange) shows strong positive contribution from X-ray luminosities, log $\lambda_\text{Edd}$, and W1-W2 colours, in contrast with negative contributions from optical emission lines and radio luminosities.}
    \label{fig:pca_coeff}
\end{figure*}

\begin{figure}
    \centering
    \includegraphics[width=1\linewidth]{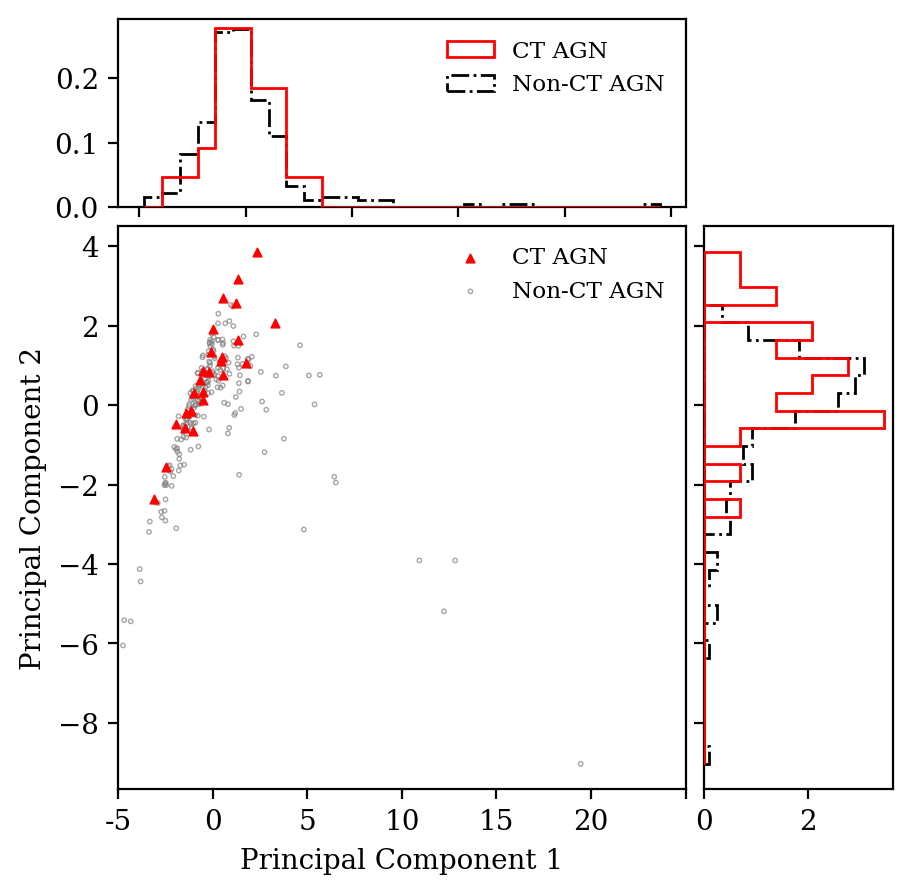}
    \caption{PCA of multi-wavelength properties for CT AGNs (red triangles) and non-CT AGNs (grey circles). The main panel shows the score plot of the first two principal components, while the histograms illustrate the normalized one-dimensional distributions of each principal component axis.}
    \label{fig:pca_plot}
\end{figure}

We combined all the multi-wavelength properties that acquired from multiple surveys to perform a multi-dimensional analysis. We adopted multi-dimensional reduction technique, Principal Component Analysis (PCA) and explored variability patterns possibly embedded in our AGN population sample, where CT AGNs might display such an instance. We have selected multi-wavelength variables such as; (i) radio luminosity ($L_\text{2-3 GHz}$) from VLASS, (ii) IR colour from WISE, (iii) optical emission line luminosities such as H$\alpha$, H$\beta$, [OIII]$\lambda$5007, [NII]$\lambda$6584, [SII]$\lambda\lambda$6716+6730, and [OI]$\lambda$6300, (iv) X-ray intrinsic luminosities such as $L_\text{10-20 keV}$, $L_\text{20-50 keV}$, and $L_\text{14-195 keV}$ (v) and Eddington ratio ($\lambda_\text{Edd}$). The percentage of variance for each principal components is shown in Figure~\ref{fig:scree_plot} while the PCA score plot for the first and second principal components are shown in Figure~\ref{fig:pca_plot}.

As shown in Figure~\ref{fig:scree_plot}, the first three principal components capture the majority of the variance in the dataset, with PC1 accounting for approximately 43\%, PC2 for 18\%, and PC3 for 11\%. Together, these components explain over 72\% of the total variance, as reflected in the characteristic `elbow' of the scree plot. Meanwhile, the loadings shown in Figure~\ref{fig:pca_coeff} provide insight into the physical representation of each component. PC1 is dominated by strong positive contributions from optical emission lines and moderate contributions from X-ray luminosities and $\lambda_\text{Edd}$, followed by minor contributions from radio luminosity and W3-W4 colours. PC2, on the other hand, shows predominantly negative contributions from optical emission lines and radio luminosities. In contrast, X-ray luminosities, $\lambda_\text{Edd}$ and W1-W2 colour display a strong positive contribution, with a small contribution from radio luminosity and longer WISE colours (W2-W3 and W3-W4).

The distribution of CT and non-CT AGNs in the PC1–PC2 plane is shown in Figure~\ref{fig:pca_plot}. Both populations overlap substantially along PC1 and PC2, with no clear offset between them. The CT AGN population is observed to be broadly disperse along the PC2 axis.

\subsection{Logistic Regression and Support Vector Machine (SVM)}
\label{sec:logistic_svm}

To investigate the multi-wavelength separation of CT and non-CT AGNs, we employed Logistic Regression and Support Vector Machines (SVM). Given the class imbalance (26 CT vs. 217 non-CT), we used the Synthetic Minority Over-sampling Technique (SMOTE; \citealt{JMLR:v18:16-365}) to balance the training set. This oversampling is crucial as it helps the classifier to identify the minority CT AGN sample, and helps mitigate the bias toward the majority class. However, given the small number of original CT sources, the results should be taken as exploratory rather than conclusive. Evaluation on an independent test set (N=49) showed robust performance, summarized in Table~\ref{tab:mlresults}. Logistic Regression achieved an accuracy of 0.84 (AUC = 0.968), while SVM reached 0.84 (AUC = 0.950). Notably, both models maintained a high recall of >0.80 for the CT class, successfully identifying four out of five CT sources.

The feature importance analysis (Table~\ref{tab:feature_importance}) demonstrates the physical drivers behind these classifications. For both models, feature importance is defined by the absolute magnitude of the learned coefficients ($|w_i|$). A higher importance value indicates a stronger influence on the classification of CT candidates. Interestingly, both models rely heavily on intrinsic X-ray luminosities. For Logistic Regression, $\log L_{\text{2-10 keV}}$ and $\log L_{\text{20-50 keV}}$ were the primary features, while the SVM shows the higher energy $\log L_{\text{20-50 keV}}$ band. Beyond X-rays, the models also provide significant insight from optical emission lines. The neutral optical line [OI]$\lambda$6300 ranked as the third most important feature in both models, followed by H$\alpha$ luminosity. In the Logistic Regression model, the $W2-W3$ colour also provided a meaningful mid-infrared diagnostic for identifying obscured candidates.


\begin{table*}
\centering
    \caption{Performance of Logistic Regression and SVM trained with SMOTE-balanced data and evaluated on the original test set.}
    \begin{tabular}{llcccccc}
    \hline
    \textbf{Model} & \textbf{Class} & \textbf{Precision} & \textbf{Recall} & \textbf{F1} & \textbf{Support} & \textbf{Accuracy} & \textbf{AUC} \\
    \hline
    Logistic Regression & CT AGN & 0.38& 1& 0.56& 5 & \multirow{2}{*}{0.84}& \multirow{2}{*}{0.968}\\
    & non-CT AGN & 1& 0.82& 0.90& 44& & \\
    \hline
    Support Vector Machine & CT AGN & 0.36& 0.80 & 0.50& 5 & \multirow{2}{*}{0.84}& \multirow{2}{*}{0.950}\\
    & non-CT AGN & 0.97& 0.84& 0.90& 44& & \\
    \hline
    \end{tabular}
    \label{tab:mlresults}
\end{table*}



\begin{table}
\centering
    \renewcommand{\arraystretch}{1.1}
    \caption{Top five features by importance for the Logistic Regression and SVM classifiers. }
    \label{tab:feature_importance}
    \begin{tabular}{c lc lc}
    \hline
    \multirow{2}{*}{\textbf{Rank}} & \multicolumn{2}{c|}{\textbf{Logistic Regression}} & \multicolumn{2}{c}{\textbf{SVM}} \\[2pt]
    \cline{2-5}
    & \textbf{Feature} & \textbf{Importance} & \textbf{Feature} & \textbf{Importance} \\
    \hline
    1 & $\log L_\text{2-10 keV}$ & 2.71 & $\log L_\text{20-50 keV}$ & 3.47 \\
    2 & $\log L_\text{20-50 keV}$ & 2.41 & $\log L_\text{2-10 keV}$ & 2.46 \\
    3 & $L_{\mathrm{[OI]\lambda6300}}$ & 1.92 & $L_{\mathrm{[OI]\lambda6300}}$  & 2.27 \\
    4 & $L_{\mathrm{H\alpha}}$ & 1.34 & $\log L_\text{14-195 keV}$ & 1.85 \\
    5 & $W2-W3$ & 0.97 & $L_{\mathrm{H\alpha}}$  & 1.43 \\
    \hline
    \end{tabular}
\end{table}


\section{Discussions}
\label{sec:discussion}

\subsection{Radio Properties and Emission Mechanism}

Overall, the consistency in the mean radio luminosity between CT and non-CT AGNs suggests that the underlying physical processes driving compact radio emission are fundamentally similar across both populations.

Interestingly, the tendency for CT AGNs to exhibit a slightly higher mean radio luminosity suggests that these sources possess robustly active radio cores. However, due to the beam size of 2.5" in VLASS, it is difficult to definitively distinguish between the underlying absorption mechanisms, such as Free-Free Absorption (FFA) or Synchrotron Self-Absorption (SSA). While FFA is commonly associated with the dense, ionized gas typical of the circumnuclear environment in obscured sources \citep{ulvestad1981radio, krolik1989physical}, SSA typically occurs within the high-brightness temperature regions of the compact core itself \citep{o1998compact}. This same dense environment might also act as a physical barrier, potentially suppressing or disrupting the formation and extension of larger-scale radio features. While these explanation remains cautious, given the beam size limitations, it will provide a plausible framework in defining why some CT AGNs often appear to posses compact radio sources, despite their high intrinsic power.

The high degree of overlap between the two populations in the radio regime reinforces the idea that the radio emission at these frequencies is largely isotropic or driven by processes—such as circumnuclear star formation or small-scale jet activity that are not strongly associated with the line-of-sight X-ray obscuration. As a result, while the CT and non-CT populations occupy distinct regions in X-ray and mid-IR colour, their radio luminosity distributions remain nearly indistinguishable, further supporting the isotropic nature of compact radio emission in these systems.

\subsection{The Radio–X-ray Connection in Obscured AGN}

The observed positive relationship between radio and X-ray luminosities across all X-ray bands, particularly in non-CT AGNs, highlights the unified relationship between accretion (X-ray emission from the innermost regions) and ejection processes (radio emission, potentially from jets or the corona).

Correlations between X-ray and radio luminosities in CT AGN and non-CT AGN samples are observed to be positive and significant, albeit a weaker one in CT AGN sample ($\rho$ $\approx$ 0.28–0.30) compared to non-CT AGN sample ($\rho$ $\geq$ 0.51). Nevertheless, the persistence of a positive correlation in CT AGNs suggests that, despite severe absorption at soft X-ray energies, their radio emission still traces non-jet-dominated X-ray radiation from the innermost accretion regions around the SMBH \citep{ballo2012exploring}. However, the observed statistical insignificance from the K-S test results between the populations across the X-ray bands, suggest that despite the overlap scatter in the radio-to-X-ray-luminosity ratio, there are parameters other than the level of obscuration that plays a role in determining the ratio. Even in higher energy bands (e.g. 20-50 keV and 14-195 keV), where the absorption is less severe, the distributions remain statistically indistinguishable. This suggests a large intrinsic scatter exists within both populations, likely driven by diverse accretion states and jet processes. This scatter dominates the distributions, effectively masking the systematic offsets observed in the regression best fits.

These results strongly support the predictions of analytical models conducted by \citet{mazzolari2024heavily}, which suggest that radio surveys offer a more complete census of the AGN population, including heavily obscured sources that X-ray observations often miss. The significantly higher surface density of CT AGNs detected in radio surveys compared to X-ray surveys, particularly at high redshift, reinforces the crucial role of radio observations in AGN studies. Future deep radio surveys, such as those conducted by the Square Kilometre Array (SKA), will refine our understanding of CT AGN demographics and their evolution, providing a more comprehensive view of AGN populations across cosmic time.

Moreover, our findings also agree with high-frequency JVLA observations reported by \citet{chiaraluce2020hard}, showing diverse radio emissions in hard X-ray-selected AGNs. The observed weak correlation between radio core (22 and 45 GHz) and 2–10 keV X-ray luminosities in their sources suggests a complex link between accretion and ejection, highlighting that radio studies are crucial for probing AGN nuclear activity.

\subsection{Mid-IR Emission and Obscuration}

The differences observed in the WISE luminosity distributions provide insight into the structure of the obscuring material. The W1 (3.4 $\mu$m) and W2 (4.6 $\mu$m) bands primarily capture emission from hot dust heated by intense radiation from the accretion disc. The lower mean luminosities observed for CT AGNs in these bands suggest that this hot dust emission is subject to heavy attenuation within the obscured environment. This aligns with the expectation that W1 and W2 may also include contributions from older stellar populations in low-luminosity AGN, where criteria such as $W1 - W2 \geq 0.8$ \citep{stern2012mid} are required to minimise host contamination.

In contrast, the W3 (12 $\mu$m) and W4 (22 $\mu$µm) bands traces warmer to cooler dust emission, predominantly from the obscuring torus at larger radial distances. The similarity in luminosity between CT and non-CT AGNs in these bands suggests that mid to far-IR emission is largely isotropic and less affected by line-of-sight absorption. In CT AGN, the absorbed  radiation is reprocessed by the torus and re-emitted at longer wavelengths \citep{pier1992infrared, netzer2015revisiting}. The W3 band includes contributions from warm dust and PAH features, while W4 isolates cooler dust emission. Both W3 and W4 appear robust against nuclear obscuration, supporting their use as reliable tracers of AGN power in heavily obscured systems.

Our findings are consistent with the behaviour of specific CT sources, such as NGC 4785 \citep{Gandhi2015}, which exhibits lower mid-IR luminosities relative to the median of the non-CT AGN population. This case study highlights that mid-IR selection alone, particularly when restricted to shorter wavelengths, may fail to identify a significant fraction of CT AGN whose IR emission is relatively weak or host-contaminated.

\subsection{Mid-IR Colours and Selection Efficiency}

The excess emission in $W1-W2$ and $W2-W3$ colours, provide us a broader understanding about the heating mechanism, mostly dominated by dust and gas surrounding the AGN, where also contamination from star-forming galaxies could possibly enhance this excess emissions.

The overlapping mean values in $W1-W2$ indicate that this colour difference alone does not provide significant separation between CT and non-CT populations. However, the broader dispersion observed in CT AGNs suggests a wider diversity in mid-IR colours. This spread is likely driven by stronger attenuation of hot-dust emission or enhanced host-galaxy contamination, which shifts a fraction of the CT AGN population to bluer colours. Consequently, WISE selection based solely on the standard $W1-W2$ cut is likely to miss a significant fraction of the CT population.

The tendency for CT AGNs to exhibit redder $W2-W3$ colours aligns with the presence of extreme dust obscuration. These redder colours are likely associated with thermal emission from a geometrically and optically thick torus surrounding the central SMBH. In this scenario, heavy obscuration attenuates the emission at shorter wavelengths, allowing cooler dust components to dominate the mid-IR spectrum \citep{gandhi2009resolving, ichikawa2017complete}.

While contamination from star-forming regions can also enhance W4 emission and mimic AGN-heated dust signatures, CT AGNs with red $W2-W3$ colours are distinct. They often display weak PAH features (e.g., at 7.7 $\mu$m), distinguishing them from starburst-dominated galaxies \citep{imanishi2010akari, alonso2014nuclear}. Thus, the $W2-W3$ excess in our CT sample is primarily a signature of the cool, obscured torus rather than host galaxy star formation.

\subsection{Reliability of WISE Colour-colour Diagram Selection}

The substantial overlap between CT and non-CT AGN in the WISE colour-colour diagram (Figure~\ref{fig:wise_diagram}) confirms that the mid-IR emission is largely isotropic. Despite the extreme line-of-sight obscuration as observed in X-ray, the reprocessed radiation from AGN-heated dust remains comparable to non-CT AGN sources. This suggest that the dusty environment, mainly the torus, is effectively re-emitting the absorbed primary radiation, thus maintaining a consistent emission across different levels of column density.

However, our bootstrap scatter analysis reveals a statistically significant difference in the dispersion between the two populations, particularly in the WISE 4-band diagram ($p=0.0001$) and the 3-band diagram ($p=0.0441$). This confirms that the observed dispersion of CT AGN WISE colours is an intrinsic physical property rather than statistical fluctuation.

As demonstrated by the fiducial density contours, the CT AGN population exhibits a distinct, non-oval and extended structure. While the 1$\sigma$ ($\sim$68\%) core density aligns with the non-CT sample, a prominent `tail' extends toward bluer $W1-W2$ and $W3-W4$ colours. This dispersion highlights the significant host galaxy contamination. In these heavily obscured sources, the primary AGN emission can be sufficiently attenuated to allow stellar emission to dilute the hot dust signature (decreasing $W1-W2$) and circumnuclear star formation to enhance cooler dust emission (increasing $W3-W4$).

For CT AGNs, the offset between the barycentre and the peak density further highlights the skewness of the population. While the average CT AGN appears redder in both $W2-W3$ and $W3-W4$, the population contains a significant minority of sources where the AGN signature may be masked by the host galaxy. The redder $W3-W4$ colours are likely associated with thermal emission from a geometrically and optically thick torus, where heavy obscuration attenuates emission at shorter wavelengths, allowing cooler dust components ($T \approx 100-300$~K) to dominate the mid-IR spectrum \citep{gandhi2009resolving, ichikawa2017complete}. 

While contamination from star-forming regions can contribute to elevated $W4$ emission, CT AGNs with red $W3-W4$ colours often display weak PAH features (e.g., at 7.7~$\mu$m), which differentiates them from starburst-dominated galaxies \citep{imanishi2010akari, alonso2014nuclear}. Our feature importance analysis (Section~\ref{sec:logistic_svm}) reflects this physical complexity. While the models mostly capture intrinsic X-ray luminosities, where it act as a indicator for CT AGN activities, the high ranking of the $W2-W3$ colour act as a key diagnostic which indicates that the classifiers are also capturing the broader SED distribution shape. This confirms that the models are effectively detecting the transition where the host galaxy's emission begins to dominate or dilute the heavily attenuated AGN signature in the mid-IR regime.

The presence of CT AGNs falling outside the \citet{mateos2012using} selection wedge suggests that mid-IR colour selection, while robust for bright and moderately obscured sources, remains incomplete for the most heavily obscured population. For these sources, multi-wavelength approaches by incorporating optical line ratios (such as $[OI]\lambda6300$) and machine learning classifiers are essential for accurate identification.


\subsection{The Physical Nature of the Narrow Line Region}

The optical emission line analysis via BPT diagram reveals the similarity between CT and non-CT AGNs, as confirmed by the bootstrap analysis and significantly overlap in the fiducial density and barycentre. The apparent scatter in the CT AGN is consistent with random sampling from the non-CT parent population. This implies that the NLR properties such as metallicity, ionization parameter, and hardness of the ionizing field are fundamentally the same for both heavily obscured and unobscured sources. This supports the Unified Model scenario where the NLR is located on scales much larger than the obscuring torus and is thus largely unaffected by the line of sight column density ($N_\text{H}$).

However, while our standard BPT diagnostics that utilise [NII]$\lambda$6584, [SII]$\lambda\lambda$6717+6730, and [OI]$\lambda$6300 provide evidence of an isotropic nature, recent studies have shown that NLR properties may vary depending on the excitation states. Specifically, \citet{bornancini2025obscured} suggested that obscured AGNs can exhibit higher ionisation parameters and lower metallicities than unobscured AGN when using specific line ratios such as [OIII]$\lambda$5007/[OII]$\lambda$3727 ($O_{32}$). For example, they found similar ionisation levels for both obscured and unobscured AGN at high-excitation states. In contrast, obscured AGN showed a lower $O_{32}$ ratio at a low-excitation state in comparison with unobscured sources. Our findings do not necessarily contradict these results, but rather suggest that the physical state of ionisation parameters may be shifting. This provides a physical context for our statistical results: the lack of statistically significant scatter in our BPT diagnostics ($p > 0.05$) indicates that the standard diagnostics may be less sensitive towards these ionisation parameters, whereas the broad dispersion observed in our mid-IR bootstrap analysis ($p = 0.0001$) aligns more closely with the heterogeneity reported by \citet{bornancini2025obscured}.

While some sources fall into LINER or star forming regions, these deviations apply equally to both CT and non-CT AGNs. For such instance, these are likely due to the sources being optically elusive, where the central engine is sufficiently obscured to remain hidden in the standard optical line diagnostic. In these cases, the observed emission line ratios is often influenced by aperture effects, where the optical spectrum includes significant light from the host galaxy (diluting the AGN emission) or could be contributions from shocks and post-AGB stars, rather than an intrinsic property that is attributed  uniquely to CT AGN.\citep{ricci2017batagn,wylezalek2020ionized}

\subsection{Accretion Properties}

For accretion properties, the negative correlation between [NII]/H$\alpha$ and $\lambda_\text{Edd}$ observed in both samples aligns with previous BASS results \citep{oh2022bass}, suggesting that higher accretion rates produce a harder ionizing continuum that suppresses low-ionization lines. The fact that CT AGNs follow this general trend, albeit with slightly more scatter, further confirms that their accretion physics is not fundamentally different from the non-CT population, despite the extreme obscuration of their local environment.

However, while the physical s linking the accretion to the ionization is consistent, the both populations are in different accretion states. The K-S test reveals a statistically distinct distribution of Eddington ratios (log $\lambda_\text{Edd}$ between the two groups ($p$=0.0019, Figure~\ref{fig:uniform_dist}). It is noteworthy to state that, the higher Eddington ratios ($\lambda_\text{Edd}$) observed among CT AGNs align well with the radiation-driven unification model \citep{ricci2017growing}. This finding is strongly supported by \citet{bornancini2025obscured} that identifies the accretion rates as the primary driver of the physical differences between obscured and unobscured AGN.

In this scenario, when accretion rates are high, the resulting powerful radiation field can blow out most of the surrounding low density gas. This means that only the densest material, specifically the high column density CT gas, can withstand this radiation pressure and remain covering the nucleus \citep{ricci2017close}. Therefore, these CT AGNs are likely observed in a phase where high obscuration and high accretion rates coexist.

\subsection{Physical Interpretation of Principal Components}

The observed patterns in Figure~\ref{fig:pca_coeff} suggest that PC1 mainly captures the total AGN output power and its subsequent impact to the surrounding environment. As AGN power increases, both high-energy X-ray and optical emission will tend to increase, which act as the primary drivers of variance for this component. However, PC2 seems to trace the balance between accretion-driven intensity and NLR emission. High PC2 values are strongly associated with the elevated X-ray luminosities, higher Eddington ratios, and redder $W1-W2$ colours, reflecting a state where accretion signatures and dust-reprocessed emission are prominent. In contrast, lower PC2 values represent environments where optical line emission is more dominant relative to the high-energy continuum.

The strong overlap between CT and non-CT AGNs in the PC1-PC2 plane (Figure~\ref{fig:pca_plot}) highlight that these component remain unaffected by the obscuring material along our line of sight. Since PC1 is linked to the AGN output power, the thick obscuring dust and gas surrounding the AGN does not change the intrinsic underlying AGN power. As a result, CT and non-CT AGN display similar AGN output power along this axis. Similarly, PC2 tracks the balance contribution from the direct and IR reprocessed emission. The overlap between both populations on this axis suggests that the presence of CT gas does not necessarily imply that a particular source possesses a more extensive dust environment or higher levels of energy reprocessing. Instead, the factors driving PC2, such as the contribution from the host galaxy and the large-scale geometry of the dust, appear to vary just as much within the CT population as they do in non-CT sources. Consequently, identifying CT AGNs remains a challenge for general photometric surveys, as their broad spectral shapes are often indistinguishable from those of less obscured, powerful AGN.

\subsection{Physical Interpretation of Machine Learning}

Although the CT AGNs sample is limited, the high recall (0.80) achieved by both the Logistic Regression and SVM models is highly encouraging. This indicates that the classifiers can reliably identify 80\% of the CT AGN population, effectively minimizing the number of missed obscured sources even under challenging conditions. The lower precision (ranging from 0.44 to 0.57) reflects some contamination from the non-CT AGN population, but does not diminish the fact that the result is physically informative. This shows that the models are picking up on a natural overlap in the data. Because the emission from the host galaxy can dilute the AGN signature, some heavily obscured sources may show similar properties to the unobscured sources.

The high ranking of intrinsic X-ray luminosities alongside optical emission lines such as [OI]$\lambda$6300 and $H\alpha$ aligns well with our PCA findings. It suggests that the models are picking up on a specific ionisation state or environmental condition, such as the presence of dense neutral gas, that is characteristic of the CT AGNs. By using SMOTE to provide a more balanced training environment, we have shown that a combination of high-energy data and specific optical emission lines can successfully isolate CT candidates even in a scenario where the sources might be fully embedded in heavily obscured. This approach offers a practical way forward for future deep surveys, allowing us to identify heavily obscured nuclei by looking at the interplay between their accretion power and their effect on the surrounding host environment.

\section{Conclusions}
\label{sec:conclusion}

In this work, we conducted a multi-wavelength analysis on Compton-thick AGNs (CT AGNs) identified from the 70-Month \textit{Swift}/BAT catalogue. By combining data from radio, IR, and optical surveys, we explored how different observational windows can offer complementary insights into these heavily obscured sources. While the sample size remains limited (26 CT AGNs and 217 Non-CT AGNs), the findings offer useful clues about the multi-wavelength characteristics of CT AGNs. Our main results are summarised as follows:

\begin{enumerate}
    \item \textbf{Radio Properties}
        \begin{enumerate}
            \item At 2-3~GHz, CT and non-CT AGNs show broadly similar luminosity distributions with both peaking near $10^{22}$ W Hz$^{-1}$.
            \item CT AGN exhibit a slightly higher mean radio luminosity compared to the non-CT sample, suggesting the presence of robustly active radio cores.
            \item While core emission may be subject to attenuation by mechanisms such as FFA or SSA, we note that the current angular resolution (e.g., from VLASS) makes it difficult to definitively distinguish between these processes.
        \end{enumerate}
    
    \item \textbf{Infrared Properties}
        \begin{enumerate}
            \item CT AGNs exhibit faint mid-IR luminosity in $W1$ and $W2$ band, suggesting IR attenuation due to heavy surrounding obscuration. In longer band ($W3$ and $W4$), CT AGN are found to be marginally more luminous, due to the cooler dust re-emission at larger radial distance, in comparison with non-CT AGNs.
            \item CT AGNs shine distinctly redder in $W2-W3$ colours, predominantly originating from cooler dust in the obscuring torus. Only a slight redder colour is observed in $W1-W2$.
            \item This dispersion observed in both $W2-W3$ and $W3-W4$ shown in the WISE colour-colour diagram are statistically significant. Our bootstrap analysis confirmed a much higher scatter in the CT population compared to non-CT AGNs ($p = 0.0441$ for 3-band and $p = 0.0001$ for 4-band data).
            \item The offset between the CT AGNs barycentre and peak density suggests that heavy obscuration often masks the AGN signature, causing many CT sources to fall outside standard WISE selection wedge.
        \end{enumerate}

    \item \textbf{Optical Diagnostics}
        \begin{enumerate}
            \item The BPT barycentres and 1$\sigma$ contours for both CT and non-CT AGNs overlap significantly, mainly occupying Seyfert regions, confirming that the NLR properties remains largely isotropic and consistent with the Unified Model.
            \item However, CT AGNs occupy a distinct accretion state, possessing higher Eddington ratios ($\lambda_{\mathrm{Edd}}$) with a significant K-S test result ($p=0.0019$).
            \item We find a negative correlation between the [NII]/H$\alpha$ ratio and Eddington ratio ($\lambda_{\mathrm{Edd}}$), indicating that more actively accreting sources tend to show stronger ionization signatures even among obscured AGNs.
            \item This higher accretion aligns with the radiation-driven unification model, suggesting that the CT state is a phase of rapid growth where intense radiation pressure allows only the densest gas to remain, while the remaining lower density gas is being pushed outward.
        \end{enumerate}
    
\item \textbf{Principal Component Analysis}  
    \begin{enumerate}
        \item The first three components capture $\sim$72\% of the variance, reducing AGN diversity to a compact parameter space.
        \item PC1 contrasts optical lines with mid-IR/X-ray luminosities, while PC2 links mid-IR, X-ray, and Eddington ratio, tracing the connection between accretion and obscuration.
        \item CT and non-CT AGNs show strong overlap in the PC1–PC2 plane, with no clear separation, indicating that PCA alone unable to distinguish between the two populations.
    \end{enumerate}

\item \textbf{Machine Learning Tests (Logistic Regression and SVM)}  
    \begin{enumerate}
        \item Logistic Regression and SVM, trained on a SMOTE-balanced set (26 CT vs 217 non-CT), achieved 0.84 accuracy (AUC: 0.95-0.97).
        \item Both models show high recall ($\geq0.80$) but low precision ($\sim$0.36--0.38), with X-ray luminosities ($L_\text{2-10 keV}$ and $L_\text{20-50 keV}$), [OI]$\lambda$6300 and H$\alpha$ luminosities, and $W2-W3$ colour as key features.
        \item The high ranking of indicators such as the [OI]$\lambda$6300 emission line and $W2-W3$ colour suggests that the models are detecting intrinsic properties of CT AGN rather than selection bias.
        \item Despite small CT numbers, results highlight the promise of multi-wavelength data, though larger samples are needed for robustness.
    \end{enumerate}

\item \textbf{Multi-wavelength Perspective}:  
    \begin{enumerate}
        \item Our results suggest that CT AGNs are not just regular AGNs that viewed from different angles, but rather, representing heterogeneous populations in a specific evolutionary states.
        \item The combination of mid-IR scatter and high Eddington ratios confirms that these sources are deeply embedded and actively growing, as supported by recent studies.
        \item This approach opens a path to robustly identifying obscured sources such as CT AGN, where the absolute luminosity may be unreliable. That said, the line ratios and multi-wavelength properties remain robust, though statistical confirmation could benefit from larger samples.
    \end{enumerate}
\end{enumerate}

Overall, our results demonstrate that a multi-wavelength approach along with a bigger CT AGN sample size is essential in identification and characterization efforts. By combining radio, IR, and optical diagnostics, we gain a more complete picture of the physical conditions and accretion processes in these deeply embedded sources beyond what X-rays alone can reveal. The PCA results indicate that the current sample size is insufficient to clearly distinguish CT from non-CT AGNs, underscoring the need for a larger dataset to establish more robust trends. Similarly, the Logistic Regression and SVM tests indicate that multi-wavelength features could potentially separate CT from non-CT AGNs, though larger datasets are required to fully establish their effectiveness.

\section*{Acknowledgements}
We acknowledge the use of the publicly available AGN catalogue from the 70-Month \textit{Swift}/BAT AGN Catalogue as our main sample for this work. Our multi-wavelength analysis made use of the Very Large Array Sky Survey (VLASS), and the AllWISE source catalogue, which is part of the Wide-field Infrared Survey Explorer (WISE) project. We thank the BAT AGN Spectroscopic Survey (BASS) collaboration for access to their Data Release 2 (DR2), which enabled the optical spectral analysis in this work.

This research was supported by the Universiti Malaya Special Grant (RUIP002-2023). We would like to extend our gratitude to the funding body for their generous support, which has been invaluable in advancing our research. Masatoshi Imanishi acknowledges the support by Japan Society for the Promotion of Science (JSPS) KAKENHI Grant Numbers 21K03632 and 25K07359.

\section*{Data Availability}

The data used in this article are publicly available from the respective survey archives and catalogues. The 70-Month \textit{Swift}/BAT AGN catalogue can be accessed from the work by \citet{ricci2017batagn} (vizier:\href{https://vizier.cds.unistra.fr/viz-bin/VizieR-3?-source=J/ApJS/233/17&-out.max=100&-out.form=HTML%20Table&-oc.form=sexa}{J/ApJS/233/17}). Radio data were obtained from the Very Large Array Sky Survey (VLASS; \href{https://public.nrao.edu/vlass/}{https://public.nrao.edu/vlass/}). Mid-IR data are available from the AllWISE source catalogue via the NASA/IPAC Infrared Science Archive (IRSA; \href{https://wise2.ipac.caltech.edu/docs/release/allwise/}{https://wise2.ipac.caltech.edu/docs/release/allwise/}). Optical spectral data were retrieved from the BAT AGN Spectroscopic Survey Data Release 2 (BASS DR2; \href{https://www.bass-survey.com}{https://www.bass-survey.com}).

Processed data products and analysis scripts used in this study are available from the corresponding author upon reasonable request.



\bibliographystyle{mnras}
\bibliography{references} 







\bsp	
\label{lastpage}
\end{document}